\documentclass[conference]{IEEEtran}
\IEEEoverridecommandlockouts
\newcommand{\bfa}{{\boldsymbol a}}

\newcommand{\bfc}{{\boldsymbol c}}

\newcommand{\bfe}{{\boldsymbol e}}

\newcommand{\bfh}{{\boldsymbol h}}

\newcommand{\diag}{\mathrm{diag}}
\newcommand{\tr}{\mathrm{tr}}

\newcommand{\bfp}{{\boldsymbol p}}

\newcommand{\bfr}{{\boldsymbol r}}

\newcommand{\bfu}{{\boldsymbol u}}

\newcommand{\bfw}{{\boldsymbol w}}
\newcommand{\bfx}{{\boldsymbol x}}

\newcommand{\bfz}{{\boldsymbol z}}
\newcommand{\bfA}{{\boldsymbol A}} %\newcommand{\bfA}{{\mathbf A}}

\newcommand{\bfC}{{\boldsymbol C}}

\newcommand{\bfJ}{{\boldsymbol J}}

\newcommand{\bfM}{{\boldsymbol M}}

\newcommand{\bfP}{{\boldsymbol P}}
\newcommand{\bfQ}{{\boldsymbol Q}}
\newcommand{\bfR}{{\boldsymbol R}}
\newcommand{\bfS}{{\boldsymbol S}}
\newcommand{\bfT}{{\boldsymbol T}}
\newcommand{\bfU}{{\boldsymbol U}}
\newcommand{\bfV}{{\boldsymbol V}}
\newcommand{\bfW}{{\boldsymbol W}}

\newcommand{\bfY}{{\boldsymbol Y}}
\newcommand{\bfZ}{{\boldsymbol Z}}

\newcommand{\rirs}{\boldsymbol{R}_{\text{RIS}}} %\newcommand{\rirs}{\mathbf{R}_{\mathrm{IRS}}}
\newcommand{\rtx}{\boldsymbol{R}_{\text{Tx}}}

\newcommand{\Phimat}{{\boldsymbol{\it{\Phi}}}} %\newcommand{\Phimat}{\mathbf{\Phi}}
\newcommand{\phivec}{\pmb{\phi}} %\newcommand{\phivec}{\pmb{\phi}}

\newcommand{\channelT}{{\sqrt{\beta}}\rirs^{1/2} {\bfW} \rtx^{1/2, \mathrm{H}}} %\newcommand{\channelT}{\sqrt{\beta}\rirs^{1/2} {\bfW} \rtx^{1/2, \mathrm{H}}}

\newcommand{\im}{\mathrm{i}}
\newcommand{\jm}{\mathrm{j}}

\newcommand{\dm}{\mathrm{d}}
\newcommand{\Tm}{\mathrm{T}}

\newcommand{\lam}{{\boldsymbol{\lambda}}}
\newcommand{\ang}{{\pmb{\varphi}}}

\newcommand{\nc}{\mathcal{N_\mathbb{C}}}

\newcommand{\Dcal}{\boldsymbol{\mathcal{D}}}
\newcommand{\Xcal}{\boldsymbol{\mathcal{X}}}

\newcommand{\Gcal}{\boldsymbol{\mathcal{G}}}
\newcommand{\Ycal}{\boldsymbol{\mathcal{Y}}}

\newcommand{\Ocal}{\mathcal{O}}

\newcommand{\Hm}{{\mathrm{H}}}

\newcommand{\Ex}{\mathop{{}\mathbb{E}}}

\newcommand{\biggamma}{\boldsymbol{\Gamma}}

\newcommand{\Tmean}{\overline{{\bfT}}}

\usepackage{cite}
\usepackage{amsmath,amssymb,amsfonts}
\usepackage{lipsum}
\usepackage{graphicx}
\usepackage{subcaption}
\usepackage{hyperref}
\usepackage{textcomp}
\usepackage{xcolor}
\def\BibTeX{{\rm B\kern-.05em{\sc i\kern-.025em b}\kern-.08em T\kern-.1667em\lower.7ex\hbox{E}\kern-.125emX}}

\usepackage{tikz}
\usepackage{pgfplots}
\pgfplotsset{compat=newest}
\pgfplotsset{plot coordinates/math parser=false}
\newlength\figH  %figureheight
\newlength\figW  %figurewidth
\usetikzlibrary{quotes,arrows.meta}
\usepackage{physics}
\usepackage{amssymb}
\usepackage{amsmath}
\usepackage{amsthm}
\usepackage{float}
\usepackage{algorithm}
\usepackage{algpseudocode}
\usepackage{algorithm,caption}
\usepackage{algorithmicx}
\usepackage[english]{babel}

\usepackage{graphicx}
\usepackage{wrapfig}
\usepackage{caption}
\usepackage{array}
\usepackage{multirow}
\usepackage{arydshln}
\usepackage{color}
\usepackage{xcolor}
\usepackage{tcolorbox}
\usepackage[framemethod=tikz]{mdframed}
\newcommand{\tikznode}[2]{%
	\ifmmode%
	\tikz[remember picture,baseline=(#1.base),inner sep=0pt] \node (#1) {$#2$};
	\else
	\tikz[remember picture,baseline=(#1.base),inner sep=0pt] \node (#1) {#2};%
	\fi}

\definecolor{darkgreen}{rgb}{0,.39,0}

\definecolor{darkred}{rgb}{0.4,0,0}

\definecolor{dnared}{rgb}{1,.27,0} 
\definecolor{dnablue}{rgb}{0,1,1}
\definecolor{dnapurple}{rgb}{0.82,0.13,0.56}
\definecolor{dnagreen}{rgb}{0,1,0}

\usepackage{afterpage}
\usepackage{tikz}
\usepackage{lipsum}

\usetikzlibrary{%
	arrows,%
	automata,%
	backgrounds,%
	calc,%
	patterns,%
	plotmarks,%
	positioning,%
	shapes.geometric,%
	shapes,%
	snakes,%
	decorations.pathmorphing,%
	decorations.shapes,%
	graphs,%
	chains,%
	arrows.meta, %
}

\usepackage{verbatim}
\usepackage{tabularx}
\usepackage{lipsum}
\usepackage{stackengine}
\usepackage{nicefrac}

\usepackage{svg}
\definecolor{bostonuniversityred}{rgb}{0.8, 0.0, 0.0}

\usepackage{xcolor}

\newcommand{\x}{\mathrm{x}}

\usepackage{mathdots}

\definecolor{lime}{HTML}{A6CE39}

\DeclareRobustCommand{\orcidicon}{%

                \begin{tikzpicture}

                \draw[lime, fill=lime] (0,0)

                circle [radius=0.16]

                node[white] {{\fontfamily{qag}\selectfont \tiny ID}};

                \draw[white, fill=white] (-0.0625,0.095)

                circle [radius=0.007];

                \end{tikzpicture}

                \hspace{-2mm}

}

\foreach \x in {A, ..., Z}{%

                \expandafter\xdef\csname orcid\x\endcsname{\noexpand\href{https://orcid.org/\csname orcidauthor\x\endcsname}{\noexpand\orcidicon}}

}

\begin{document}

\title{Design of a Multi-User RIS-Aided System with Statistical Channel Knowledge}

\author{\IEEEauthorblockN{Sadaf~Syed \orcidA{}, ~\IEEEmembership{Graduate Student~Member,~IEEE}, Dominik~Semmler \orcidB{}, ~\IEEEmembership{Graduate Student~Member,~IEEE}, \\ Donia Ben Amor \orcidC{}, ~\IEEEmembership{Graduate Student~Member,~IEEE}, Michael Joham \orcidD{}, ~\IEEEmembership{Member,~IEEE}, \\ and Wolfgang Utschick \orcidE{}, ~\IEEEmembership{Fellow,~IEEE} }
\IEEEauthorblockA{{School of Computation, Information and Technology, Technical University of Munich, Germany}\\
Emails: \{sadaf.syed, dominik.semmler, donia.ben-amor, joham, utschick\}@tum.de}

}

\maketitle

\begin{abstract}
Reconfigurable intelligent surface~(RIS) is a promising technology to enhance the spectral and energy efficiency in a wireless communication system. The design of the phase shifts of an RIS in every channel coherence interval demands a huge training overhead, making its deployment practically infeasible. The design complexity can be significantly reduced by exploiting the second-order statistics of the channels. This paper is the extension of our previous work to the design of an RIS for the multi-user setup, where we employ maximisation of the lower bound of the achievable sum-rate of the users. Unlike for the single-user case, obtaining a closed-form expression for the update of the filters and phase shifts is more challenging in the multi-user case. We resort to the fractional programming~(FP) approach and the non-convex block coordinate descent~(BCD) method to solve the optimisation problem. As the phase shifts of the RIS obtained by the proposed algorithms are based on the statistical channel knowledge, they do not need to be updated in every channel coherence interval.  
%Furthermore, we do not consider any Gaussian approximation of the effective RIS channel like our previous work, which makes the lower bound more involved.    
\end{abstract}

\section{Introduction}
\IEEEPARstart{M}{assive} multiple-input multiple-output (MaMIMO) systems, wherein the base station (BS) is equipped with a huge number of antennas, have the capacity to fulfill the growing need for high data throughput and reduced energy consumption in the wireless communication systems~\cite{lu2014overview}. Nevertheless, deploying a large number of antennas would increase the circuit energy consumption and the hardware expenditures. The recently proposed technology of reconfigurable intelligent surface~(RIS) has emerged as an alternative cost-effective solution to enhance the spectral efficiency~\cite{wu2019towards, huang2019reconfigurable, renzo2019smart}. Since the advent of wireless communication systems, the propagation medium between the transmitter and the receiver has been perceived as an uncontrollable entity. The RIS has been proposed as a solution to overcome the ill effects of the propagation medium by manipulating the direction of the transmitted signal to the user in a controlled fashion. It is composed of a large number of passive reflecting elements and offers the flexibility to configure these passive elements in order to achieve the desired performance objectives. Moreover, the passive nature of the RIS elements eliminates the need for an active transmit radio frequency (RF) chain, thereby resulting in a significantly reduced energy consumption and hardware cost compared to the traditional active antennas at the BS. Thus, they can be scaled economically compared to the scaling of active antennas at the BS.
\par In this paper, we investigate an RIS-aided multi-user multiple-input single-output~(MISO) downlink~(DL) communication system and try to exploit the second-order channel statistics in order to reduce the system's design complexity.
\subsection{Related Works}
The design of the RIS for various system models has been extensively investigated in the literature, and it offers a considerable performance gain in comparison to the systems without an RIS \cite{wu2019intelligent, guo2020weighted, zhang2020capacity, wu2018intelligent, relying, semmler2022linear, syed2023design, hu2020statistical, twotime, dang2020joint, gan2021ris, zhang2022sum, noh2022cell}. For example, the transmit power of the BS is minimised by employing an RIS in \cite{wu2018intelligent,wu2019intelligent}, where an alternating optimisation approach is proposed, which relies on the semi-definite relaxation (SDR) technique. Although this algorithm provides a good performance, its complexity is very high, especially for a large-sized RIS. The algorithm proposed in \cite{guo2020weighted} aims to maximise the weighted sum-rate of the users by employing an iterative approach, which utilises the fractional programming (FP) technique \cite{fractional1, fractional2}. \cite{relying} also considers maximisation of the weighted sum-rate of the users by employing the Majorisation-Minimisation algorithm \cite{sun2016majorization} and the Complex Circle Manifold method \cite{circle}. However, in order to fully exploit the potential of the RIS, the algorithms in \cite{wu2019intelligent, guo2020weighted, zhang2020capacity, wu2018intelligent, relying, semmler2022linear} require perfect knowledge of the RIS-associated links in every channel coherence interval. The passive architecture of the RIS intricates the channel estimation of its links to the BS and the users, and the requirement to estimate these channels perfectly in every channel coherence interval would lead to a huge training overhead, e.g., \cite{joham2022estimation}. Moreover, these algorithms demand the joint optimisation of the phase shifts and the transmit filters to be performed in every channel coherence interval, which is computationally very expensive for an RIS-aided system with a large number of passive elements, and also practically infeasible to be implemented for the fast-fading channels.     
\par In order to combat this problem, many algorithms have been proposed that exploit the statistical knowledge of the channels to design the phase shifts of the RIS~\cite{syed2023design, hu2020statistical, twotime, dang2020joint, gan2021ris, zhang2022sum, noh2022cell}. Since the structure of the channels varies slowly, the channel statistics remain constant for many channel coherence intervals, and hence, it is relatively easier to obtain accurate information of the second-order statistics of the channels through long-term observation. The phase shifts designed on the basis of the channel covariance matrices do not need to be updated regularly, i.e., there is no need to perform the joint optimisation in every channel coherence interval. Moreover, the RIS-associated channels do not need to be estimated frequently as in the case of the algorithms in~\cite{wu2019intelligent, guo2020weighted, zhang2020capacity, wu2018intelligent, relying, semmler2022linear}. This significantly reduces the channel training overhead and the design complexity of the RIS.
\par The algorithms in \cite{syed2023design, hu2020statistical, dang2020joint} propose low complexity methods for a single-user RIS-aided system based on the long-term channel statistics, but these algorithms cannot be readily extended to the multi-user case. The algorithms in \cite{gan2021ris, zhang2022sum} consider a multi-user system, but they assume specific channel models for analytical simplicity. For example, \cite{gan2021ris} considers a Rician fading model for the RIS-associated channels, which assumes a deterministic line-of-sight~(LoS) component and the non-line-of-sight~(NLoS) component to have independent and identically distributed~(i.i.d.) entries following complex Gaussian distribution with zero mean and unit variance. Similarly, \cite{zhang2022sum} assumes a system model with no direct path and the Kronecker channel model \cite{kermoal2002stochastic} with zero mean for all the channels. In these algorithms, the phase shifts of the RIS and the transmit filters at the BS are only optimised in the coherence interval of the covariance matrices. The algorithms in \cite{twotime}, \cite{noh2022cell} and \cite{statCSI} propose a two-timescale~(TTS) approach with hybrid offline and online optimisation phases. In the offline phase, the phase shifts of the RIS are designed considering the long-term statistics of the channels, which are then kept fixed in the coherence interval of the covariance matrices. On the other hand, in the online step, the optimal filters are computed in every channel coherence interval using the perfect knowledge of the instantaneous channel state information~(CSI). This leads to a better performance than the algorithms in \cite{gan2021ris, zhang2022sum} but at the cost of a higher computational complexity.  
\subsection{Contributions}
\par Motivated by the above challenges, we propose algorithms for designing a multi-user RIS-aided system considering a more general channel model, where the computational complexity is reduced by exploiting the statistical knowledge of the channels. The main contributions of this paper can be summarised as follows.
\begin{itemize}
    \item We consider a general channel model where the direct channel between the BS and the user and the channel between the RIS and the user are considered to be zero mean with perfectly known covariance matrices. Since the RIS is expected to be deployed at a favourable position, we assume a deterministic LoS path between the BS and the RIS in addition to a NLoS component. This system model is different from the one considered in \cite{gan2021ris} and \cite{noh2022cell}, where a deterministic LoS path for all the channels is assumed. The proposed algorithms are based on the maximisation of the sum-rate of the users by employing the information-theoretic lower bound on the achievable rate~\cite{medard} as the figure of merit.   
    \item The Gaussian assumption of the effective channel in an RIS-aided system holds only asymptotically for an infinite number of reflecting elements~\cite{wang2022massive}. Unlike our previous work \cite{syed2023design}, where we considered the asymptotic Gaussian behaviour to derive the closed-form expression of the lower bound on the sum-rate~\cite{medard}, we derive the lower bound for the non-Gaussian channel model. 
    \item We first propose an algorithm where both the transmit filter at the BS and the phase shifts of the RIS need to be optimised only in the coherence interval of the covariance matrices. A generalised matched filter~(GMF) or the bilinear precoder \cite{amor2020bilinear, gmf} is used as the transmit filter at the BS.
    %The algorithm has very low complexity as the entire optimisation procedure needs to be performed less frequently. However, this algorithm is observed to offer a good performance only for the single-user case and at the low transmit power levels for the multi-user scenario since the bilinear precoder is unable to mitigate the inter-user interference in the multi-user case, and as such, a saturation effect is observed in the high transmit power regime. % for the multi-user setup. 
    \item  The algorithm based on the GMF is observed to have a saturation of the sum-rate of the users in the high transmit power regime for the multi-user setup. In order to combat this problem, we also propose another algorithm that employs the TTS approach of \cite{twotime}, where the phase shifts of the RIS are optimised in the coherence interval of the channel statistics, whereas the optimal filters are designed in every channel coherence interval. Both the algorithms require the channel to be estimated in every channel coherence interval, however, we only need to estimate the effective channel of the system and not the RIS-associated links separately. We compare our TTS algorithm to the one in \cite{twotime}, which is based on the stochastic successive convex approximation~(SSCA) method \cite{liu2018online} employing the maximisation of a concave surrogate function during the offline step. %, in which several random channel samples are generated from their known distributions in each iteration. % For each of the generated samples, the optimal filters are computed by employing the iterative weighted minimum mean-square error~(IWMMSE) method \cite{wmmse} and this process needs to be repeated several times till the convergence is reached. The computation of the filters with the IWMMSE method makes the algorithm computationally demanding since the IWMMSE method itself is computationally expensive. 
\end{itemize}  
\subsection{Organisation of the Paper and Mathematical Notation}
The paper is organised as follows. In Section~II, we present the system model and formulate the optimisation problem. In Section~III, we present a simplification of the problem to a tractable form. Section~IV deals with the optimisation procedure and the simulation results are presented in Section~V. Finally, we conclude the work in Section~VI. 
\par \textbf{Notation}: Scalars, vectors and matrices are respectively denoted by lower/upper case, boldface lower case and boldface upper case letters. For an arbitrary matrix $\bfA$, $\bfA^{\Tm}$ and $\bfA^{\Hm}$ represent its transpose and conjugate transpose respectively. $\big[\bfA\big]_{\im,\jm}$ signifies the element in the $i$-th row and the $j$-th column of the matrix $\bfA$. The expectation and the variance of a random variable are defined by $\Ex[.]$ and var(.) respectively, while $\tr(.)$ denotes the trace of a square matrix. The Hadamard product of two matrices of the same dimensions is represented by $\odot$ operator, whereas $\otimes$ symbolises the Kronecker product of two matrices of arbitrary dimensions. The real and the imaginary parts of a complex number are defined by $\Re{.}$ and $\Im{.}$ respectively. The operators $(.)^{*}$, $|.|$ and $\|.\|$ stand for the complex conjugation, scalar magnitude and $\mathcal{L}$-2 norm respectively, whereas $\text{vec}(.)$ denotes vectorisation of the input matrix by stacking its columns into a vector.  
\section{System Model and Problem Formulation}
%In this work, the system model considered is similar to our previous work in \cite{syed2023design}, except that we have multiple single-antenna users now. 
The DL of an RIS-aided multi-user MISO communication system is investigated. The system is equipped with one BS having $M$ antennas and serving $K$ single-antenna users. The RIS consists of $N$ passive reflecting elements. The phase-shift matrix is defined by the diagonal matrix $\Phimat$ = $\diag$($\phi_1,\cdots,\phi_N$), where $\phi_1 \cdots \phi_N$ are the phase shifts of the $N$ elements of the RIS and $\phivec = [\phi_1, \cdots, \phi_N]^{\Tm}$ denotes the phase-shift vector. The channel from the BS to the $k$-th user is denoted by $\bfh_{\dm, k}\in \mathbb{C}^{M}$ and it is circularly symmetric, complex Gaussian distributed, with zero mean and covariance matrix $\bfC_{\dm,k}$, i.e., $\bfh_{\dm,k}\sim\mathcal{N_\mathbb{C}}({\bf{0}}, \bfC_{\dm,k})$. The channel from the RIS to the user is denoted by $\bfr_{k}\in \mathbb{C}^{N}$ with $\bfr_{k}\sim\mathcal{N_\mathbb{C}}({\bf{0}}, \bfC_{\bfr, k})$. The channel from the BS to the RIS is assumed to have a deterministic LoS component in addition to the random part, which is modelled as
\begin{align}
{\bfT} = \underbrace{\left(\sqrt{1 - \beta}\right)\bfT'}_{\bar{\bfT}} + \channelT \label{eqn1}
\end{align}
where $\left(\sqrt{1 - \beta}\right)\bfT' = \bar{\bfT}$ denotes the LoS component between the BS and the RIS. The entries in $\bfW\in \mathbb{C}^{N \times M}$ are i.i.d. with unit variance and zero mean, following circularly symmetric complex Gaussian distributions. $\bfR_{\text{RIS}}$ and $\bfR_{\text{Tx}}$ denote the channel correlation matrices on the side of the RIS and the BS respectively, and $\beta \in [0,1]$ is the factor controlling the strength of the LoS component. Note that the extreme case of $\beta = 1$ corresponds to the case of zero-mean $\bfT$, similar to that in \cite{syed2023design}.    
The effective channel for the $k$-th user is given by
\begin{align}
\label{eqn4.1}
\bfh_{k} & = \bfh_{\dm,k} + \bfT^{\Hm}\Phimat\bfr_{k} 
\end{align}
where $\bfh_{\dm,k}, \:\bfT$ and $\bfr_{k}$ are mutually independent. Note that $\bfh_k$ is zero-mean, and its covariance matrix is given by $\bfC_k$. Moreover, it is assumed that the channels of different users are mutually independent, i.e., $\Ex[\bfh_i\bfh_j^{\Hm}] = {\mathbf{0}} \:\forall\:i\neq j $. It is clear from~\eqref{eqn4.1} that $\bfh_k$ is non-Gaussian since $\bfT$ and $\bfr_k$ follow Gaussian distributions. The signal received by the $k$-th user is 
\begin{align}
\label{eqn4.2}
 y_{k} = \bfh_{k}^{\Hm}\bfp_{k}\:s_{k} + \sum\limits_{i \neq k}\bfh_{k}^{\Hm}\bfp_{i}\:s_{i} + v_{k}
\end{align}
where $\bfp_{k}\in \mathbb{C}^{M \times 1}$ is the transmit beamforming vector for the $k$-th user, $s_{k}\!\sim\mathcal{N_\mathbb{C}}(0, 1)$ is the transmitted data symbol to user $k$ and $v_{k} \sim\mathcal{N_\mathbb{C}}(0, 1)$ denotes the additive white Gaussian noise~(AWGN) at the $k$-th user's side. The objective is to reduce the training complexity, and hence, to update the phase shifts of the RIS only in each coherence interval of the covariance matrices, which is much longer than the channel coherence interval. The sum-rate optimisation problem with respect to the transmit filters and the phase shifts of the RIS reads as
\begin{align}
&\!\max\limits_{\bfp_1, \cdots, \bfp_K, \phivec}      &\qquad&  \mathop{{}\mathbb{E}}\left[\sum\limits_{k=1}^{K}\log_2\left(1 + \gamma_k \right) \right] \nonumber \\
& \quad \text{s.t.}  &     &  \gamma_k  = \frac{\big|{\bfh_{k}^{\Hm}}{\bfp_{k}}\big|^2}{\sum\limits_{\substack{j = 1 \\ j \neq k}}^{K}\big|{\bfh_{k}^{\Hm}}{\bfp_{j}}\big|^2 + 1} \label{eqn8b} \tag{P1}\\
&     &      &  \sum\limits_{k=1}^{K}\mathop{{}\mathbb{E}}[\|{\bfp_k}\|^2] \leq P \nonumber \\
&                  &      & |\phi_n| = 1 \:\forall \:n = 1, \cdots, N.  \nonumber
\end{align}
Here $\gamma_k$ denotes the $k$-th user's signal-to-interference-plus-noise-ratio (SINR) and $P$ denotes the total transmit power budget of the BS. It is very difficult to obtain a closed-form expression of the expected sum-rate of the users, and hence, we employ a lower bound on the sum-rate based on the worst-case noise, which is extensively used in the massive MIMO literature~\cite{medard}. The lower bound of the users' sum-rate is given by $\sum\limits_{k=1}^{K}\log_{2}(1 + \gamma_k^{\mathrm{lb}})$, where $\gamma_k^{\mathrm{lb}}$ is the lower bound of the actual SINR of the $k$-th user, expressed as
\begin{align}
\label{lb}
\gamma_k^{\mathrm{lb}} &= \dfrac{\left|\mathop{{}\mathbb{E}}\left[\bfh_k^{\Hm}\bfp_k\right]\right|^2}{\mathrm{var}({\bfh_k^{\Hm}\bfp_k})  + \sum\limits_{\substack{j = 1 \\ j \neq k}}^{K}{\mathop{{}\mathbb{E}}\left[\left|{{\bfh_k^{\Hm}}{\bfp_j}}\right|^2\right] + 1}}.
\end{align}
In order to obtain the closed-form expressions of the expectation and the variance terms in the lower bound, we choose $\bfp_k$ to be a GMF, also known as the bilinear precoder \cite{amor2020bilinear, neumann2017generalized}, given by   
\begin{align}
\label{precoder}
\bfp_k = \bfA_k \bfh_k
\end{align} 
where $\bfA_k \in \mathbb{C}^{M\times M}$ is a deterministic transformation matrix, which only depends on the higher-order statistical knowledge of the channels. As such, the optimisation problem in \eqref{eqn8b} is modified to the maximisation of the lower bound, given by
\begin{align}
&\!\max\limits_{\bfA_1, \cdots, \bfA_K, \phivec}      &\qquad&  \sum\limits_{k=1}^{K}\log_2\left(1 + \gamma_k^{\mathrm{lb}} \right)  \label{eqnP2} \tag{P2} \\
& \quad \text{s.t.}  &     & \sum\limits_{k=1}^{K}\mathop{{}\mathbb{E}}[\|{\bfp_k}\|^2] = \sum\limits_{k=1}^{K}\tr(\bfA_k \bfC_k \bfA_k^{H}) \leq P \nonumber \\
&                  &      & |\phi_n| = 1 \:\forall \:n = 1, \cdots, N. \nonumber
\end{align}
\section{Simplification of the Optimisation Problem}
In order to solve \eqref{eqnP2}, we need to express the optimisation problem in a closed-form. Since $\bfh_k$ is not Gaussian distributed, the variance term in \eqref{lb} is no longer similar to the one derived in \cite{amor2020bilinear, neumann2017generalized}.  \\
{\it{\bf{Theorem 1}}}: For $\bfp_k = \bfA_k\bfh_k$ and $\bfh_k$ given by \eqref{eqn4.1}, $\mathrm{var}({\bfh_k^{\Hm}\bfp_k}) = \bfa_k^{\Hm}\bfJ_k\bfa_k + \tr({\bfA_k \bfC_k \bfA_k^{\Hm}\bfC_k})$, where
\begin{align*}
     \bfJ_k &=   \beta^2 \tr\left(\bfQ\rirs \right) \left(\text{vec}(\rtx)\text{vec}^{\Hm}(\rtx)  + (\rtx^{\Tm}\otimes \rtx) \right) \\
     &+ 2\beta \Re{\text{vec}(\rtx) \text{vec}^{\Hm}(\bar{\bfT}^{\Hm}\bfQ \bar{\bfT})}  + \beta \rtx^{\Tm}\otimes \left( \bar{\bfT}^{\Hm}\bfQ\bar{\bfT}\right) \\
      & + \beta\left( \bar{\bfT}^{\Hm}\bfQ \bar{\bfT}\right)^{\Tm} \otimes \rtx \\
      \bfQ &= \Phimat \bfC_{\bfr,k}\Phimat^{\Hm}\rirs \Phimat \bfC_{\bfr,k}\Phimat^{\Hm}.
 \end{align*}
\begin{proof}
Please refer to Appendix B.
\end{proof}
Using \eqref{precoder}, we get $\mathop{{}\mathbb{E}}\left[\bfh_k^{\Hm}\bfp_k\right] = \tr(\bfC_k\bfA_k)$ and $\mathop{{}\mathbb{E}}\left[\left|{{\bfh_k^{\Hm}}{\bfp_j}}\right|^2\right] = \tr(\bfC_k\bfA_j\bfC_j\bfA_j^{\Hm})$. Thus, the lower bound of the SINR in~\eqref{lb} can be expressed in a closed-form as 
 \begin{align}  
\gamma_k^{\mathrm{lb}} %& = \dfrac{\left|{\bfc_k^{\Hm}\bfa_k}\right|^2}{\sum\limits_{j=1}^{K}{\tr\left( \bfC_k \bfA_j\bfC_j \bfA_j^{\Hm}\right)} + \bfa_k^{\Hm}\bfJ_k\bfa_k + 1}  \nonumber\\
& = \dfrac{\left|{\bfc_k^{\Hm}\bfa_k}\right|^2}{\sum\limits_{j=1}^{K}{\bfa_j^{\Hm}\left(\bfC_j^{\Tm} \otimes \bfC_k \right) \bfa_j} + \bfa_k^{\Hm}\bfJ_k\bfa_k + 1} \label{lb2}
 \end{align}
 where $\bfa_k$ = $\text{vec}(\bfA_k)$ and $\bfc_k = \text{vec}(\bfC_k)$. The covariance matrix $\bfC_k$ depends on the phase shifts of the RIS, and hence, $\bfC_k$ must be expressed as a function of $\phivec$ for the phase shift optimisation. 
\subsection{Computation of the Channel Covariance Matrix}
The channel covariance matrix of the effective channel for the $k$-th user can be computed as
\begin{align}
 &\bfC_k  = \mathop{{}\mathbb{E}}\left[\bfh_k\bfh_k^{\Hm}\right]  = {\bfC_{\dm, k}}  +  \mathop{{}\mathbb{E}_{\bfr_k, \bfT}} \left[\bfT^{\Hm}\Phimat \bfr_k \bfr_k^{\Hm}\Phimat^{\Hm} \bfT\right] \nonumber \\
&\stackrel{(a)}{=} \bfC_{\dm,k} + \Tmean^{\Hm}\Phimat\bfC_{\bfr,k}\Phimat^{\Hm}\Tmean \nonumber \\ & + \beta \mathop{{}\mathbb{E}_{\bfW}} \left[\rtx^{1/2} \bfW^{\mathrm{H}}\rirs^{1/2, \mathrm{H}}\Phimat \bfC_{\bfr,k} \Phimat^{\mathrm{H}}\rirs^{1/2} \bfW \rtx^{1/2, \mathrm{H}}\right] \nonumber
 \end{align}
where ($a$) follows from \eqref{eqn1} and using the fact that $\bfh_{\dm,k}$, $\bfr_k$ and $\bfW$ are mutually independent random variables. Since the entries of $\bfW$ are i.i.d. with zero mean and unit variance, and $\Phimat = \diag(\phivec)$, the above expression can be simplified as
 \begin{align}
   &\bfC_k  = \bfC_{\dm,k} + \Tmean^{\Hm}\Phimat\bfC_{\bfr,k}\Phimat^{\Hm}\Tmean + \beta {\mathrm{tr}}(\rirs {\Phimat} \bfC_{\bfr,k} {\Phimat}^{\mathrm{H}})\rtx \nonumber \\
  &= \bfC_{\dm,k} + \Tmean^{\Hm}\Phimat\bfC_{\bfr,k}\Phimat^{\Hm}\Tmean + \beta {\mathrm{tr}}\Big(\rirs (\bfC_{\bfr,k} \odot \phivec \phivec^{\mathrm{H}})\Big)\rtx  \nonumber \\
   &\stackrel{(b)}{=} \bfC_{\dm,k} + \Tmean^{\Hm}\Phimat\bfC_{\bfr,k}\Phimat^{\Hm}\Tmean + \beta \phivec^{\mathrm{H}} \left(\rirs \odot \bfC_{\bfr, k}^{\mathrm{T}}\right)\phivec \rtx \label{eqnC}
 \end{align} 
where ($b$) uses Lemma~1 of \cite{syed2023design}.  
\subsection{Reformulation of the Objective Function}
With the expression of $\gamma_k^{\mathrm{lb}}$ in \eqref{lb2}, the objective function in \eqref{eqnP2} becomes very involved. \eqref{eqnP2} can be simplified by the FP approach \cite{fractional1, fractional2} in a more tractable form. Introducing the auxiliary variables $ \lambda_k$, the objective function in \eqref{eqnP2} can be reformulated as 
\begin{align}
    \sum\limits_{k=1}^{K}\log(1 + \gamma_k^{\mathrm{lb}}) = \: \!\max\limits_{\lambda_k \geq 0} \: \sum\limits_{k=1}^{K}\log(1 + \lambda_k) - \lambda_k + \dfrac{(1 + \lambda_k)\gamma_k^{\mathrm{lb}}}{1 + \gamma_k^{\mathrm{lb}}}. \label{fp}
\end{align}
The last fractional term in \eqref{fp} is given by
\begin{align}
   &\dfrac{(1 + \lambda_k)\gamma_k^{\mathrm{lb}}}{1 + \gamma_k^{\mathrm{lb}}}\\ & = \dfrac{(1 + \lambda_k)\left|{\bfc_k^{\Hm}\bfa_k}\right|^2}{\left|{\bfc_k^{\Hm}\bfa_k}\right|^2 + \sum\limits_{j=1}^{K}{\bfa_j^{\Hm}\left(\bfC_j^{\Tm} \otimes \bfC_k \right) \bfa_j} + \bfa_k^{\Hm}\bfJ_k \bfa_k   + 1}. \nonumber
\end{align}
Using the quadratic transform step of the FP approach, the numerator and the denominator of the above fraction can be decoupled with the auxiliary variables $\chi_k$ as
\begin{align}
\begin{split}
   &\dfrac{(1 + \lambda_k)\gamma_k^{\mathrm{lb}}}{1 + \gamma_k^{\mathrm{lb}}}  = 2 \text{Re}\left \{\chi_k^{*}\left(\sqrt{1 + \lambda_k}\right)\bfc_k^{\Hm}\bfa_k \right\} \\
   &  - \left|\chi_k\right|^2\left(\left|{\bfc_k^{\Hm}\bfa_k}\right|^2 + \sum\limits_{j=1}^{K}{\bfa_j^{\Hm}\left(\bfC_j^{\Tm} \otimes \bfC_k \right) \bfa_j} + \bfa_k^{\Hm}\bfJ_k \bfa_k+ 1 \right). \nonumber
   \end{split}
\end{align}
Let $\bfa = [\bfa_1, \cdots, \bfa_K]^{\Tm}$, $\lam = [\lambda_1, \cdots, \lambda_K]^{\Tm}$ and ${\pmb{\chi}}~=~[\chi_1, \cdots, \chi_k]^{\Tm}$, the objective problem can now be expressed in a more tractable form as
\begin{align}
    \begin{split}
   & \quad \quad \!\max\limits_{\bfa, \phivec, \lam, {\pmb{\chi}} } \:\:  \:     f(\bfa, \phivec, \lam, {\pmb{\chi}}), \: \text{where} \\
     &f(\bfa, \phivec, \lam, {\pmb{\chi}})  = \sum\limits_{k=1}^{K}\Big(\log\left(1 + \lambda_k \right) - \lambda_k \Big) \\ & + \sum\limits_{k=1}^{K} 2\left(\sqrt{1 + \lambda_k}\right)\text{Re}\left\{\chi_k^{*}\bfc_k^{\Hm}\bfa_k\right\}  \\& - \sum\limits_{k=1}^{K}\left|\chi_k\right|^2\left(\left|{\bfc_k^{\Hm}\bfa_k}\right|^2 + \sum\limits_{j=1}^{K}{\bfa_j^{\Hm}\left(\bfC_j^{\Tm} \otimes \bfC_k \right) \bfa_j} + \bfa_k^{\Hm}\bfJ_k \bfa_k + 1 \right) \nonumber
    \end{split}    
\end{align}
\begin{align}
    &\text{s.t.} &      &  \sum\limits_{k=1}^{K}\mathop{{}\mathbb{E}}[\|{\bfp_k}\|^2] = \sum\limits_{k=1}^{K}{{}\mathbb{E}}[\|\bfA_k \bfh_k \|^2] \leq P \nonumber\\
&                  &      & |\phi_n| = 1 \:\forall \:n = 1, \cdots, N \label{eqnP3} \tag{P3} \\
 &                 &      & \lambda_k \geq 0 \:\forall \:k = 1, \cdots, K \nonumber.
\end{align}

The optimisation problem \eqref{eqnP3} is non-convex and one can adopt the iterative non-convex BCD method \cite{bcd, bcdmain} to decompose the problem into four disjoint blocks, containing one of the four variables as the optimisation parameter in each block, as done in \cite{guo2020weighted}. The solution to \eqref{eqnP3} with the non-convex BCD method is discussed in the next section.
\section{Optimisation Steps}
The objective problem in \eqref{eqnP3} can be solved by updating the variables $\bfa$, $\phivec$, $\lam$ and ${\pmb{\chi}}$ iteratively using the BCD method. Denoting the optimisation results from the previous iteration by $\bar{\bfa}$, $\bar{\phivec}$, $\bar{\lam}$ and $\bar{{\pmb{\chi}}}$, the update rule of the variables is described next.
\subsection{Update of the auxiliary variables}
 \text{It} can be observed from \eqref{fp} that the optimal $\lambda_k$ for a fixed $\gamma_k^{\mathrm{lb}}$ is given by
\begin{align}
    \lambda_k = \gamma_k^{\mathrm{lb}}. \label{lamk}
\end{align}

For the update of ${\pmb{\chi}}$, $f(\bfa, \phivec, \lam, {\pmb{\chi}})$ has to be maximised w.r.t. ${\pmb{\chi}}$, keeping the other variables fixed. Ignoring the constant terms, the optimisation problem now becomes
\begin{align}
\label{5.26}
   & \!\max\limits_{{\pmb{\chi}}} \:\:  \:   \sum\limits_{k=1}^{K}  2\left(\sqrt{1 + \bar{\lambda}_k}\right)\text{Re}\left\{\chi_k^{*}\bar{\bfc}_k^{\Hm}\bar{\bfa}_k\right\} \nonumber \\ &- \sum\limits_{k=1}^{K}\left|\chi_k\right|^2\Bigg(\left|{\bar{\bfc}_k^{\Hm}\bar{\bfa}_k}\right|^2 + \sum\limits_{j=1}^{K}{\bar{\bfa}_j^{\Hm}\left(\bar{\bfC}_j^{\Tm} \otimes \bar{\bfC}_k \right) \bar{\bfa}_j}  \nonumber \\ & \quad \quad \quad \quad +\bfa_k^{\Hm}\bfJ_k \bfa_k + 1 \Bigg). 
\end{align}
The closed-form solution to the above problem is
\begin{align}
\chi_k &= \dfrac{\Big(\sqrt{1 + \bar{\lambda}_k}\Big)\bar{\bfc}_k^{\Hm}\bar{\bfa}_k}{\left|{\bar{\bfc}_k^{\Hm}\bar{\bfa}_k}\right|^2 + \sum\limits_{j=1}^{K}{\bar{\bfa}_j^{\Hm}\left(\bar{\bfC}_j^{\Tm} \otimes \bar{\bfC}_k \right) \bar{\bfa}_j} +\bfa_k^{\Hm}\bfJ_k \bfa_k  + 1}. \label{5.28} 
\end{align}

\subsection{Update of the transformation matrix}
The optimisation problem for this block reads as
\begin{align}
    \begin{split}
    &\!\max\limits_{\bfa_1, \cdots, \bfa_K} \:\:  \:     \sum\limits_{k=1}^{K} 2\left(\sqrt{1 + \bar{\lambda}_k}\right)\text{Re}\left\{\bar{\chi}_k^{*}\bar{\bfc}_k^{\Hm}\bfa_k\right\} \\ & - \sum\limits_{k=1}^{K}\left|\bar{\chi}_k\right|^2\Bigg(\left|{\bar{\bfc}_k^{\Hm}\bfa_k}\right|^2 + \sum\limits_{j=1}^{K}{\bfa_j^{\Hm}\left(\bar{\bfC}_j^{\Tm} \otimes \bar{\bfC}_k \right) \bfa_j} \\& \quad \quad \quad +\bfa_k^{\Hm}\bfJ_k \bfa_k  + 1 \Bigg)
    \end{split}    
\end{align}
\begin{align}
    & \quad \text{s.t.} &      & \sum\limits_{k=1}^{K}{{}\mathbb{E}}[\|\bfA_k\bfh_k \|^2] =  \sum\limits_{k=1}^{K} \tr\left( \bfA_k \bar{\bfC}_k \bfA_k^{\Hm}\right) \leq P. \label{5.30}
\end{align}
The constraint in \eqref{5.30} can be equivalently represented in the vectorised form as
\begin{align}
   \sum\limits_{k=1}^{K} \tr\left( \bfA_k \bar{\bfC}_k \bfA_k^{\Hm}\right) =   \sum\limits_{k=1}^{K} \bfa_k^{\Hm}\left(\bar{\bfC}_k^{\Tm}\otimes {\bf{I}}\right)\bfa_k\leq P. \label{5.31}
\end{align}
The optimal solution to the above optimisation problem is given by 
\begin{align}
   \bfa_k &=  \bar{\chi}_k\left(\sqrt{1 + \bar{\lambda}_k}\right)\Bigg(\left|\bar{\chi}_k\right|^2 \bar{\bfc}_k \bar{\bfc}_k^{\Hm} + \left|\bar{\chi}_k\right|^2 \bfJ_k  \nonumber \\ &+ \sum\limits_{j=1}^{K} \left|\bar{\chi}_j\right|^2\Big(\bar{\bfC}_k^{\Tm}\otimes \bar{\bfC}_j\Big) + \mu\Big(\bar{\bfC}_k^{\Tm}\otimes {\bf{I}} \Big)\Bigg)^{-1}\bar{\bfc}_k\label{5.32}
\end{align}
where $\mu \geq 0$ is the optimal Lagrangian multiplier corresponding to the DL power constraint in~\eqref{5.31}, which can be obtained using one-dimensional search techniques, e.g., the bisection method. However, the matrix inversion is expensive and one needs to perform a large number of iterations to obtain an accurate value of $\mu$ by the bisection method. It can become computationally very expensive if $M$ is large. This problem can be averted by applying the trick mentioned in \cite{joham2002transmit, christensen2008weighted, zhao2023rethinking}. The idea is to convert the constrained optimisation problem in~\eqref{5.30} to an equivalent unconstrained problem by introducing the constraint into the objective function itself. The inequality constraint in \eqref{5.31} will be satisfied with equality for the optimal precoding filters, and hence, \eqref{5.30} can be written as
\begin{align}
    \begin{split}
    &\!\max\limits_{\bfa_1, \cdots, \bfa_K} \:\:  \:     \sum\limits_{k=1}^{K} 2\left(\sqrt{1 + \bar{\lambda}_k}\right)\text{Re}\left\{\bar{\chi}_k^{*}\bar{\bfc}_k^{\Hm}\bfa_k\right\} \\ & - \sum\limits_{k=1}^{K}\left|\bar{\chi}_k\right|^2\Bigg(\left|{\bar{\bfc}_k^{\Hm}\bfa_k}\right|^2 + \sum\limits_{j=1}^{K}{\bfa_j^{\Hm}\left(\bar{\bfC}_j^{\Tm} \otimes \bar{\bfC}_k \right) \bfa_j} \\ & \quad \quad \quad +\bfa_k^{\Hm}\bfJ_k \bfa_k  +  \dfrac{\sum\limits_{j=1}^{K} \bfa_j^{\Hm}\left(\bar{\bfC}_j^{\Tm}\otimes {\bf{I}}\right)\bfa_j}{P} \Bigg). \label{eqn16}
   \end{split}    
\end{align}
The optimal solution of \eqref{eqn16} is given by
    \begin{align}
   &\bfa_k =  \bar{\chi}_k \left(\sqrt{1 +  \bar{\lambda}_k}\right)\Bigg(\left|\bar{\chi}_k\right|^2 \bar{\bfc}_k \bar{\bfc}_k^{\Hm} + \left|\bar{\chi}_k\right|^2 \bfJ_k  \nonumber \\ &+ \sum\limits_{j=1}^{K} \left|\bar{\chi}_j\right|^2\Big(\bar{\bfC}_k^{\Tm}\otimes \bar{\bfC}_j\Big)  + \sum\limits_{j=1}^{K}\dfrac{\left|\bar{\chi}_j\right|^2}{P}\Big(\bar{\bfC}_k^{\Tm}\otimes {\bf{I}} \Big)\Bigg)^{-1}\bar{\bfc}_k. \label{5.33}
\end{align}
 Hence, we obtain a closed-form solution for the optimal precoders without the need to compute $\mu$ by the iterative one-dimensional search methods. However, it is to be noted that the optimal $\bfa_k$ from \eqref{5.33} might not necessarily satisfy the downlink power constraint of \eqref{5.31}. Hence, $\bfa_k$ from \eqref{5.33} must be scaled in the end to fulfill the power budget.

\subsection{Update of the phase-shift vector}
Ignoring the constant terms, the optimisation problem for this block is given by
\begin{align}
    \begin{split}
    &\!\max\limits_{\phivec} \:\:  \:     \sum\limits_{k=1}^{K} 2\left(\sqrt{1 + \bar{\lambda}_k}\right)\text{Re}\left\{\bar{\chi}_k^{*}{\bfc}_k^{\Hm}\bar{\bfa}_k\right\} \\ & - \sum\limits_{k=1}^{K}\left|\bar{\chi}_k\right|^2\Biggl(\left|{{\bfc}_k^{\Hm}\bar{\bfa}_k}\right|^2 + \sum\limits_{j=1}^{K}{\bar{\bfa}_j^{\Hm}\left({\bfC}_j^{\Tm} \otimes {\bfC}_k \right) \bar{\bfa}_j} 
 +\bfa_k^{\Hm}\bfJ_k \bfa_k\Biggr) \label{6.20} 
      \end{split}    
\end{align}
\begin{align}
  \hspace{-45pt} & \text{s.t.} & |\phi_n| = 1 \:\forall \:n = 1, \cdots, N. \label{6.21}
\end{align}
$\bfC_k$ from \eqref{eqnC} is inserted in \eqref{6.20}, and $\phivec$ is replaced by $e^{j \ang}$, where $\phi_n~= ~e^{j \varphi_n} \: \:\forall\:n$ and $\ang \in \mathbb{R}^{N \times 1}$ denotes the vector containing the corresponding angles in radians. We would then employ the iterative gradient ascent method to find the optimal $\ang$. The gradient of \eqref{6.20} w.r.t. $\ang$ is given by $ 2 \:\text{Re}\big\{ -j \bar{\phivec^{*}} \odot {\boldsymbol{\Delta}} \big\}$, where the expression of $\boldsymbol{\Delta}$ is provided in Appendix~C.
Hence, the update rule for $\ang$ can be written as
\begin{align} \label{6.22}
\ang \leftarrow \bar{\ang} + 2\:\kappa\:\text{Re}\big\{ -j \bar{\phivec^{*}} \odot {\boldsymbol{\Delta}} \big\}
\end{align}
where $\kappa$ is the optimal step size, which can be computed by the Armijo rule \cite{armijo}. 

The iterative BCD method results in the improvement of the objective function in each sub iteration and its convergence has been established in \cite{bcd, bcdmain}. Once the optimal phase shift vector $\phivec$ and the transformation matrices $\bfA_k$ are obtained, they are kept fixed in the coherence interval of the covariance matrices. The filters are then updated in each channel coherence interval by \eqref{precoder}, i.e., the filter update in each channel coherence interval only requires multiplication of the precomputed matrices $\bfA_k$ with the perfectly estimated channel $\bfh_k$. The proposed algorithm is summarised in Algorithm~1. 
\begin{algorithm}
\caption{Bilinear Precoders based Design of RIS using the Statistical Channel Knowledge}
\begin{algorithmic}[1]
    \State  Initialise the phase-shift vector $\phivec$ and the auxiliary variables $\lam$ and ${\pmb{\chi}}$ with feasible values;
    \State  Initialise $\bfa$ according to \eqref{5.33};
    \State {\textbf{Repeat}}
        \State Update $\lam$, ${\pmb{\chi}}$ and $\bfa$ according to \eqref{lamk}, \eqref{5.28}, and \eqref{5.33} respectively;
        \State Find the optimal step size $\kappa$ by the Armijo rule and update $\ang$ according to \eqref{6.22};
    \State \textbf{Until} The value of the objective function in \eqref{eqnP3} converges.  
    \State Obtain the precoding vectors by \eqref{precoder}. 
\end{algorithmic}
\end{algorithm}

It will be later shown in the simulation results that the proposed algorithm offers a performance gain with low complexity. However, it is also observed that the GMFs cannot suppress the inter-user interference, which results in a saturation of the sum-rate at high power levels \cite{neumann2017generalized, neumann2018bilinear, amor2020bilinear}. This leads to a considerable performance loss compared to the algorithms requiring instantaneous CSI for the phase shift update~\cite{guo2020weighted} at high transmit power. This saturation problem can be resolved by employing a hybrid online and offline optimisation algorithm as adopted in \cite{statCSI, twotime,noh2022cell}. The key idea here is to design the phase shifts based on the channels' statistical knowledge so that we do not need to compute the phases in every channel coherence interval. The phase shift optimisation is again performed by maximising the lower bound of the users' sum rate given in \eqref{eqnP3}. Once the optimised phase shifts are obtained, they are kept fixed in the coherence interval of the covariance matrices. The design of the phase shifts belongs to the offline optimisation part, which relies only on the long-term statistics of the channels. After performing the offline optimisation, the transmit filters are now designed in the online optimisation step, i.e., the filter optimisation~(for fixed phase shifts) is now performed in every channel coherence interval using the BCD or zero-forcing~(ZF) methods, instead of multiplying the perfectly estimated channel vector with the precomputed matrices as in Algorithm~1. After estimating the effective channel vectors for the users, the filter update step in Algorithm~1 has the complexity of $\Ocal(M^2)$, but this hybrid method computing the optimal filters in every channel coherence interval has the complexity of $\Ocal(M^3)$. However, the filter design in every channel coherence interval still does not require the knowledge of the RIS-associated links separately. Instead, the BS only requires the knowledge of the effective channels, thereby reducing the channel training overhead as compared to the algorithms only based on the instantaneous CSI, e.g.,~\cite{guo2020weighted}. The overall hybrid algorithm is summarised in Algorithm~2.
\begin{algorithm}
\caption{Hybrid Online and Offline Design of RIS using the Statistical Channel Knowledge}
\begin{algorithmic}[1]   
    \State Perform the offline optimisation with the statistical knowledge of the channels:
      \begin{itemize}
        \item  Initialise the phase-shift vector $\phivec$ and the auxiliary variables $\lam$ and ${\pmb{\chi}}$ with feasible values;
        \item Initialise $\bfa$ according to \eqref{5.33};
        \item {\bf{Repeat}}
        \begin{itemize}
            \item  Update $\lam$, ${\pmb{\chi}}$ and $\bfa$ according to \eqref{lamk}, \eqref{5.28}, and \eqref{5.33} respectively;
            \item  Find the optimal step size $\kappa$ by the Armijo rule and update $\ang$ according to \eqref{6.22};
        \end{itemize}
        \item {\bf{Until}} The value of the objective function in \eqref{eqnP3} converges.
    \end{itemize}
       \State The phase-shift vector obtained from the offline optimisation step is kept fixed and the online optimisation step is performed in each channel coherence interval:
       \begin{itemize}
           \item Apply either the iterative BCD algorithm or the zero-forcing method with the optimal power allocation (waterfilling solution \cite{telatar1999capacity}) to obtain the precoding vectors $\bfp_1, \cdots, \bfp_K$. 
       \end{itemize}
\end{algorithmic}
\end{algorithm}
\subsection{Convergence and Complexity}
\begin{figure}		
		 \scalebox{1} {\definecolor{mycolor1}{rgb}{1.00000,0.00000,1.00000}%
\definecolor{mycolor2}{rgb}{0.00000,1.00000,1.00000}%
\definecolor{mycolor3}{rgb}{0.92900,0.69400,0.12500}%
\definecolor{mycolor4}{rgb}{0.49400,0.18400,0.55600}%
\definecolor{mycolor5}{rgb}{0.46600,0.67400,0.18800}%
\definecolor{mycolor6}{rgb}{0.30100,0.74500,0.93300}%
\definecolor{mycolor7}{rgb}{0.63500,0.07800,0.18400}%
\begin{tikzpicture}

\begin{axis}[%
width=0.9\figW,
height=\figH,
at={(0\figW,0\figH)},
scale only axis,
xmin=0,
xmax=250,
xlabel style={font=\color{white!15!black}},
xlabel={$\bf{Iteration\:No.}$},
ymin=0,
ymax=12,
ylabel style={font=\color{white!15!black}},
ylabel={$\bf{Sum-Rate\:[bpcu]}$},
axis background/.style={fill=white},
xmajorgrids,
ymajorgrids,
legend style={at={(0.95,0.2)}, anchor=south east, legend cell align=left, align=left, draw=white!15!black, row sep=-0.05cm}
]

\addplot [color=mycolor3, line width=1.0pt, mark=diamond, mark options={solid, mycolor3}]
   table[row sep=crcr]{%
1	0.926221041319400\\
2	1.258447000145496\\
3	1.578039361942033\\
4	1.733207620909953\\
5	1.815504535584000\\
6	1.861132270917112\\
7   1.887023198620768\\
8   1.901919397651981\\
9   1.910557951897721\\
10  1.915591198121206\\
11  1.918532342955971\\
12  1.920254274893245\\
13  1.921263829045159\\
};
\addlegendentry{${P = -10\:\text{dB}}$}

\addplot [color=mycolor1, line width=1.0pt, mark=diamond, mark options={solid, mycolor1}]
   table[row sep=crcr]{%
1	1.44874456316824\\
2	1.70390631214675\\
3	1.97707863473889\\
4	2.25468405235951\\
5	2.48074205583764\\
6	2.64513793383487\\
7	2.75878106688142\\
8	2.83929481680545\\
9	2.89967265819604\\
10	2.94866480064861\\
11	2.99059616230918\\
12	3.02835309417301\\
13	3.06325103232919\\
14	3.09645101493465\\
15	3.12849840160009\\
16	3.15998763203678\\
17	3.19118196219804\\
18	3.22238203695427\\
19	3.25367134005679\\
20	3.28514392670982\\
21	3.31674575587549\\
22	3.34842560980364\\
23	3.38004184442976\\
24	3.41146508707683\\
25	3.44252441610004\\
26	3.47307471776882\\
27	3.50296328843078\\
28	3.5320705361615\\
29	3.56028663575769\\
30	3.58753398156187\\
31	3.61374975264123\\
32	3.63889741598043\\
33	3.66295374377771\\
34	3.68591446821897\\
35	3.70778493213062\\
36	3.72858287964924\\
37	3.74833196451014\\
38	3.76706321352151\\
39	3.78481065275032\\
40	3.80161216815928\\
41	3.81750662591716\\
42	3.83253445159163\\
43	3.84673576836955\\
44	3.8601508323213\\
45	3.87281884811575\\
46	3.88477831886505\\
47	3.89606630586768\\
48	3.90671871913264\\
49	3.91676986505566\\
50	3.92625269086373\\
51	3.9351985174411\\
52	3.94363724586838\\
53	3.95159720832313\\
54	3.95910534275836\\
55	3.96618711833844\\
56	3.97286668321638\\
57	3.97916683634202\\
58	3.98510915256519\\
59	3.99071398265758\\
60	3.99600055936296\\
61	4.00098701397057\\
62	4.00569046639511\\
63	4.0101270508577\\
64	4.01431199259509\\
65	4.01825963794196\\
66	4.02198351984547\\
67	4.02549638941838\\
68	4.02881027206904\\
69	4.03193649874557\\
70	4.03488575418203\\
71	4.03766810679502\\
72	4.04029305028887\\
73	4.04276953164156\\
74	4.04510598709699\\
75	4.04731036797915\\
76	4.04939017192326\\
77	4.05135246644461\\
78	4.0532039161164\\
79	4.05495080393333\\
80	4.0565990550252\\
81	4.05815425592203\\
82	4.05962167529527\\
83	4.06100628126629\\
84	4.06231275958815\\
85	4.06354552915383\\
86	4.0647087579672\\
87	4.06580637701088\\
88	4.06684209430051\\
};
\addlegendentry{${P = 0\:\text{dB}}$}
\addplot [color=mycolor2, line width=1.0pt, mark=o, mark options={solid, mycolor2}]
  table[row sep=crcr]{%
1	1.53824749386041\\
2	1.85168610725064\\
3	2.2428589699855\\
4	2.79354516827556\\
5	3.51951020667531\\
6	4.26282628508967\\
7	5.00411962531007\\
8	5.63682231895786\\
9	6.11393461132097\\
10	6.49191316443226\\
11	6.80283630191823\\
12	7.06546826147988\\
13	7.29173645544657\\
14	7.48968053174749\\
15	7.66497034452242\\
16	7.82175053466444\\
17	7.96314046670169\\
18	8.09154528902771\\
19	8.20885759978589\\
20	8.31659278071892\\
21	8.41598252520475\\
22	8.50804113289143\\
23	8.59361354922407\\
24	8.67341085855418\\
25	8.74803696221681\\
26	8.81800894079264\\
27	8.88377281126842\\
28	8.94571587311943\\
29	9.00417649147317\\
30	9.05945192950595\\
31	9.11180467833648\\
32	9.16146761705556\\
33	9.20864825275122\\
34	9.25353223032527\\
35	9.29628625777917\\
36	9.33706055985949\\
37	9.37599094833391\\
38	9.41320057849386\\
39	9.44880144718618\\
40	9.48289567663961\\
41	9.51557661975798\\
42	9.54692981581812\\
43	9.57703382018487\\
44	9.60596092742191\\
45	9.63377780379218\\
46	9.6605460424109\\
47	9.68632265210541\\
48	9.71116048923864\\
49	9.73510864027976\\
50	9.75821276169575\\
51	9.78051538273698\\
52	9.80205617585978\\
53	9.822872198837\\
54	9.84299811203024\\
55	9.86246637380815\\
56	9.88130741668792\\
57	9.89954980642899\\
58	9.91722038601172\\
59	9.93434440618728\\
60	9.95094564406584\\
61	9.96704651102799\\
62	9.98266815108757\\
63	9.99783053069737\\
64	10.0125525208687\\
65	10.0268519723787\\
66	10.0407457847462\\
67	10.0542499695819\\
68	10.0673797088517\\
69	10.0801494085315\\
70	10.0925727480802\\
71	10.1046627261147\\
72	10.116431702628\\
73	10.1278914380551\\
74	10.1390531294665\\
75	10.1499274441328\\
76	10.1605245506869\\
77	10.1708541480833\\
78	10.1809254925385\\
79	10.1907474226163\\
80	10.2003283826071\\
81	10.2096764443385\\
82	10.2187993275396\\
83	10.22770441887\\
84	10.2363987897183\\
85	10.244889212861\\
86	10.2531821780685\\
87	10.261283906735\\
88	10.2692003656043\\
89	10.2769372796567\\
90	10.2845001442159\\
91	10.2918942363327\\
92	10.2991246254948\\
93	10.3061961837091\\
94	10.3131135949996\\
95	10.3198813643618\\
96	10.3265038262067\\
97	10.332985152332\\
98	10.3393293594479\\
99	10.3455403162897\\
100	10.3516217503409\\
101	10.3575772541926\\
102	10.3634102915636\\
103	10.3691242029998\\
104	10.3747222112735\\
105	10.3802074265026\\
106	10.3855828510043\\
107	10.3908513839014\\
108	10.3960158254926\\
109	10.4010788814056\\
110	10.4060431665415\\
111	10.4109112088242\\
112	10.415685452766\\
113	10.4203682628598\\
114	10.4249619268077\\
115	10.429468658595\\
116	10.4338906014174\\
117	10.4382298304718\\
118	10.4424883556154\\
119	10.4466681239031\\
120	10.4507710220049\\
121	10.4547988785171\\
122	10.4587534661649\\
123	10.4626365039087\\
124	10.4664496589547\\
125	10.4701945486772\\
126	10.4738727424553\\
127	10.4774857634306\\
128	10.4810350901878\\
129	10.4845221583632\\
130	10.4879483621837\\
131	10.4913150559417\\
132	10.4946235554071\\
133	10.4978751391788\\
134	10.5010710499832\\
135	10.5042124959146\\
136	10.5073006516297\\
137	10.5103366594885\\
138	10.5133216306518\\
139	10.5162566461349\\
140	10.5191427578164\\
141	10.5219809894093\\
142	10.5247723373939\\
143	10.5275177719114\\
144	10.5302182376256\\
145	10.5328746545499\\
146	10.5354879188415\\
147	10.5380589035676\\
148	10.5405884594402\\
149	10.5430774155238\\
150	10.5455265799168\\
151	10.5479367404078\\
152	10.5503086651054\\
153	10.5526431030477\\
154	10.5549407847876\\
155	10.5572024229563\\
156	10.5594287128078\\
157	10.5616203327437\\
158	10.5637779448162\\
159	10.5659021952183\\
160	10.5679937147516\\
161	10.5700531192814\\
162	10.572081010172\\
163	10.574077974712\\
164	10.5760445865185\\
165	10.5779814059334\\
166	10.5798889804016\\
167	10.581767844839\\
168	10.5836185219864\\
169	10.5854415227527\\
170	10.5872373465457\\
171	10.5890064815914\\
172	10.5907494052456\\
173	10.5924665842895\\
174	10.5941584752222\\
175	10.5958255245398\\
176	10.5974681690044\\
177	10.5990868359079\\
178	10.6006819433257\\
179	10.6022539003606\\
180	10.603803107381\\
181	10.6053299562518\\
182	10.6068348305559\\
183	10.6083181058116\\
184	10.6097801496798\\
185	10.6112213221678\\
186	10.6126419758252\\
187	10.6140424559341\\
188	10.6154231006945\\
189	10.6167842414012\\
190	10.6181262026189\\
191	10.6194493023497\\
192	10.6207538521967\\
193	10.622040157521\\
194	10.623308517597\\
195	10.6245592257605\\
196	10.6257925695528\\
197	10.6270088308623\\
198	10.6282082860601\\
199	10.6293912061322\\
200	10.630557856809\\
201	10.6317084986886\\
202	10.6328433873598\\
203	10.6339627735185\\
204	10.6350669030826\\
205	10.6361560173048\\
206	10.6372303528778\\
207	10.6382901420417\\
208	10.639335612686\\
209	10.6403669884481\\
210	10.6413844888103\\
211	10.6423883291941\\
};
\addlegendentry{${P = 30\:\text{dB}}$}
\end{axis}

\end{tikzpicture}%
}
		\caption{Convergence Plot for IWMMSE method for $K = 3$, $M = 8$}
		\label{wmmse_conv}
\end{figure}
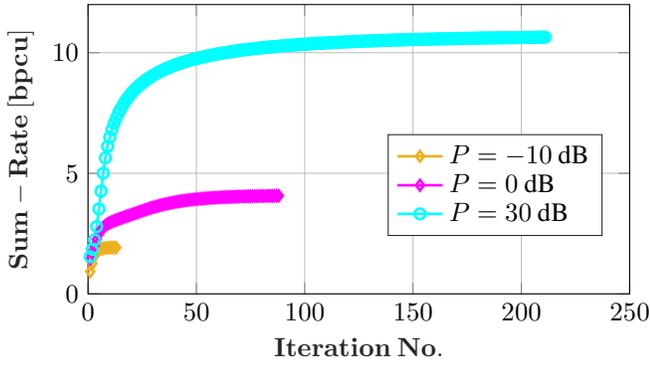

The convergence of the BCD algorithm is discussed in \cite{bcdmain}. Since the value of the objective function improves in each sub-problem, the convergence to a local minimum is guaranteed. The reduction in the complexity by the proposed algorithms is achieved due to the fact that the phase shifts of the RIS only need to be updated in the coherence interval of the covariance matrices. Both of the algorithms employ the same update step for the phase shifts of the RIS. The complexity for updating the phase shifts by the proposed algorithms is $\Ocal(I_a(KN^2 + KNM^2))$, where $I_a$ is the number of iterations required by the algorithms for convergence. We next compare the complexity of the proposed algorithms with the TTS algorithm in~\cite{twotime}, which uses the SSCA method in which several random channel samples are generated from their known distributions in each iteration. For each of the generated samples, the optimal filters are computed by employing the iterative weighted minimum mean-square error~(IWMMSE) method \cite{wmmse} and this process needs to be repeated several times till the convergence is reached. %The computation of the filters with the IWMMSE method makes the algorithm computationally demanding since the IWMMSE method itself is computationally expensive. 
The optimisation of the phase shifts with the SSCA based TTS method~\cite{twotime} has the complexity of $\Ocal(I_b(KNM + T_{H}JKM^3))$, where $I_b$ denotes the number of iterations required for convergence, $J$ denotes the number of IWMMSE iterations, and $T_{H}$ denotes the number of random channel samples required for the SSCA method. Even though the complexity of our proposed algorithms grows with $N^2$, whereas the SSCA method has a linear growth with $N$, using Algorithms 1 and 2 results in a reduced computational time for the practical systems. This is due to the fact that the complexity of the SSCA method is dominated by the requirement to perform IWMMSE iterations several times. From the convergence plot of the IWMMSE method in Fig.~\ref{wmmse_conv}, it can be seen that its convergence is very slow for high power levels, and it takes more than 100 iterations to converge for $P = 30$~dB. Each IWMMSE iteration in Fig.~\ref{wmmse_conv} needs several iterations to compute the optimal Lagrangian parameter, which is needed for the update of the transmit filters in \cite{wmmse}~(discussed in \cite{zhang2023discerning}). However, the proposed algorithms bypass this requirement because of the closed-form solution given by \eqref{5.33}. Moreover, the IWMMSE method needs to be applied for each of the $T_H$ random samples, which is generally around 50 to 100 in number. This increases the computational time of the TTS approach in \cite{twotime} and its run-time is observed to be approximately 10 times higher than that of the propsoed algorithms.          
\section{Simulation Results}
\begin{figure}	
    \centering
   \scalebox{0.7} {\begin{tikzpicture}[node distance=2cm]
       \fill[violet] (-4,-2) -- node[below, black] {BS (0 m, 0 m)}  (-3,-2) -- (-3.5,-1) -- cycle ;
       % \node[circle, black, draw, label=below:User (0 m)](a) at (2.5,-1.8){};
       \draw (2.5,-1.8) circle (1.0cm); 
        \draw (2.0,-1.4) circle (0.2cm) node[label=right:$k\text{-th user}$]{};
         \draw (3.0,-2.0) circle (0.2cm);
         \draw (2.0,-2.2) circle (0.2cm);
       \draw[step=0.2cm,black,thin, xshift=0cm,yshift=0cm] (-1,-0.6) grid (0.8,0.6)
       node[label=above:$\text{RIS ({50} m, 10 m)}$]{};
        \draw[-{Stealth[length=2.5mm, width=1.5mm]},dashed] (-3.5,-1.8) -- (2.0,-1.4) node[midway, above]{$\bfh_{\dm, k}$};
        \draw (-3.5,-1.8) -- (2.4,-1.8) node[midway, below]{$\text{{\it{D}} m}$};
         \draw (2.4,-1.6) -- (2.4,-2.0);
        \draw[-{Stealth[length=2.5mm, width=1.5mm]},dashed] (-3.5,-1.8) -- (-1,0.10) node[midway,sloped, above]{$\bfT$};
        \draw[-{Stealth[length=2.5mm, width=1.5mm]},dashed] (0.9,0.0) -- (2.0,-1.4) node[midway,sloped, above]{$\bfr_k$};
         \draw[-{Stealth[length=3mm, width=2mm]}] (-3.5,-1.8) -- (-3.5,0.7);
         \draw[-{Stealth[length=3mm, width=2mm]}] (3.5,-1.8) -- (4.2,-1.8);
    \end{tikzpicture}}
    \caption{Simulation Setup}
    \label{setup1}
\end{figure}
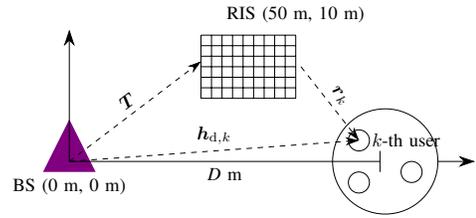

In this section, numerical results are provided to validate the effectiveness of the proposed algorithms. The system comprises of one BS equipped with $M = 16$ antennas, serving $K$ single-antenna users. The RIS is equipped with $N = 100$ passive reflecting elements. The setup is illustrated in Fig.~\ref{setup1}. The users are placed inside a circle of radius $50\,\text{m}$ with the centre of the circle at a distance $D\,\text{m}$ from the BS. Each of the channels is generated according to its distribution described in Section~II. The covariance matrix of each channel is generated according to the 3GPP specifications \cite{etsi5138}. The BS, users and the RIS are equipped with a uniform linear array (ULA). To model each of the channels, a spatial channel propagation model is considered similar to \cite{amor2023rate}, where the channels are obtained by the superposition of multiple-path components resulting from the scatterers in the surrounding propagation environment. Assuming $N_{\text{path}} = 6$ main clusters and $N_{\text{ray}} = 20$ rays in each cluster, each covariance matrix is given by 
\begin{align*}
    \bfC_{i,k} = \alpha_{i,k}\sum\limits_{n=1}^{N_\text{path}} \dfrac{\nu_{n}}{N_\text{ray}}\sum\limits_{m=1}^{N_\text{ray}}\bfx_{i,k}(\theta_{n,m})\bfx_{i,k}^{\Hm}(\theta_{n,m})
\end{align*}
where $\theta_{n,m}$ corresponds to the incident angle of the received signal from the scatterer of the $m$-th ray in the $n$-th cluster, and the index $i$ is used to differentiate the different components of the effective channel, i.e., $\bfh_{\dm,k}$, $\bfr_k$ and $\bfT$. $\bfx_{i,k}(\theta_{n,m})$ denotes the array steering vector of the $i$-th channel component of the $k$-th user at an angle~$\theta_{n,m}$ and is given by 
\begin{align*}
\bfx_{i,k}(\theta_{n,m}) = \left[1, \mathrm{e}^{j\pi \mathrm{sin}(\theta_{n,m})}, \cdots, \mathrm{e}^{j\pi(M_i - 1)\mathrm{sin}(\theta_{n,m})}\right]^{\Tm}
\end{align*}
where $M_i$ denotes the number of antennas. $\nu_{n}$ signifies the power of the $n$-th cluster and $\alpha_{i,k}$ corresponds to the slow-fading coefficient which depends on the distance. According to the 3GPP model, $\alpha_{i,k}$ is equal to the signal to noise ratio~(SNR) at distance $\delta_{i,k}$ and is given by \cite{bjornson2016massive}  
\begin{align*}
10\log_{10}(\alpha_{i,k}) = 78.7 - 37.6 \log_{10}(\delta_{i,k}) \:\mathrm{dB}.
\end{align*}
The deterministic LoS component $\bfT'$ is considered to have rank one, and it is generated by taking the outer product of two array-steering vectors of lengths $N$ and $M$ respectively, and $\beta$ is set to 0.2. The achievable sum-rate of the users is taken as the performance metric, which is averaged over 1000 channel realisations for 100 different covariance matrices $\bfC_{i,k}$, generated by varying the position of the users and the path powers $\nu_n$. 
 
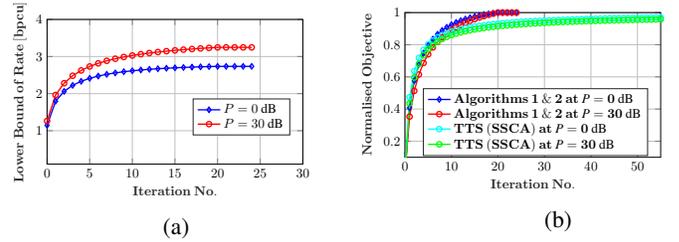
\begin{figure}		
\begin{minipage}{.26\textwidth}
		\scalebox{0.5}{% This file was created by matlab2tikz.
\definecolor{mycolor1}{rgb}{0.00000,0.44700,0.74100}%
\definecolor{mycolor2}{rgb}{0.85000,0.32500,0.09800}%
\begin{tikzpicture}

\begin{axis}[%
width=0.85\figW,
height=\figH,
at={(0\figW,0\figH)},
scale only axis,
xmin=0,
xmax=30,
xlabel style={font=\color{white!15!black}},
xlabel={$\bf{Iteration\:No.}$},
ymin=0.1,
ymax=4,
ylabel style={font=\color{white!15!black}},
ylabel={$\bf{Lower\:Bound\:of\:Rate\:[bpcu]}$},
axis background/.style={fill=white},
xmajorgrids,
ymajorgrids,
legend style={at={(0.95,0.2)}, anchor=south east, legend cell align=left, align=left, draw=white!15!black, row sep=-0.05cm}
]
\addplot [color=blue, line width=1.0pt, mark=diamond, mark options={solid, blue}]
  table[row sep=crcr]{%
0	1.14691977714937\\
1	1.79540915546524\\
2	2.06187505069189\\
3	2.22298417251015\\
4	2.33269010562668\\
5	2.4125001013445\\
6	2.47312875831463\\
7	2.52062666745437\\
8	2.55870694548502\\
9	2.58978962377625\\
10	2.61552621938563\\
11	2.63708569499829\\
12	2.65532046819351\\
13	2.67086774358215\\
14	2.68421396169743\\
15	2.69573722435591\\
16	2.70573605256526\\
17	2.71444937903233\\
18	2.72207075750925\\
19	2.7287586605369\\
20	2.73464407305886\\
21	2.73464407305886\\
22	2.73464407305886\\
23	2.73464407305886\\
24	2.73464407305886\\
};
\addlegendentry{${P = 0\:\text{dB}}$}

\addplot [color=red, line width=1.0pt, mark=o, mark options={solid, red}]
  table[row sep=crcr]{%
0	1.26355116167284\\
1	1.96334066131033\\
2	2.28089799703911\\
3	2.48227761921529\\
4	2.62478162297823\\
5	2.73205754094761\\
6	2.81620081507056\\
7	2.88419210160768\\
8	2.94039517140828\\
9	2.98769972888187\\
10	3.02810528989257\\
11	3.06304438311456\\
12	3.09357284735868\\
13	3.1204874849746\\
14	3.14440177200579\\
15	3.16579624553114\\
16	3.18505299775271\\
17	3.20247985956346\\
18	3.21832775410118\\
19	3.2328033764292\\
20	3.24607857885354\\
21	3.24607857885354\\
22	3.24607857885354\\
23	3.24607857885354\\
24	3.24607857885354\\
};
\addlegendentry{${P = 30\:\text{dB}}$}
h\end{axis}

\begin{axis}[%
width=1.227\figW,
height=1.227\figH,
at={(-0.16\figW,-0.135\figH)},
scale only axis,
xmin=0,
xmax=1,
ymin=0,
ymax=1,
axis line style={draw=none},
ticks=none,
axis x line*=bottom,
axis y line*=left
]
\end{axis}
\end{tikzpicture}%
} 
		\subcaption{}
		\label{convergence_K1}
  \end{minipage}%
  \begin{minipage}{.3\textwidth}
  \scalebox{0.5}{% This file was created by matlab2tikz.
\definecolor{mycolor1}{rgb}{0.00000,0.44700,0.74100}%
\definecolor{mycolor2}{rgb}{0.85000,0.32500,0.09800}%
\definecolor{aqua}{rgb}{0.0, 1.0, 1.0}
\begin{tikzpicture}

\begin{axis}[%
width=0.85\figW,
height=\figH,
at={(0\figW,0\figH)},
scale only axis,
xmin=0,
xmax=55,
xlabel style={font=\color{white!15!black}},
xlabel={$\bf{Iteration\:No.}$},
ymin=0.1,
ymax=1,
ylabel style={font=\color{white!15!black}},
ylabel={$\bf{Normalised\:Objective}$},
axis background/.style={fill=white},
xmajorgrids,
ymajorgrids,
legend style={at={(0.95,0.01)}, anchor=south east, legend cell align=left, align=left, draw=white!15!black, row sep=-0.05cm}
]
\addplot [color=blue, line width=1.0pt, mark=diamond, mark options={solid, blue}]
  table[row sep=crcr]{%
0    0\\
1    0.4084\\
2    0.5763\\
3    0.6777\\
4    0.7468\\
5    0.7971\\
6    0.8353\\
7    0.8652\\
8    0.8892\\
9    0.9088\\
10    0.9250\\
11    0.9386\\
12    0.9500\\
13    0.9598\\
14    0.9682\\
15    0.9755\\
16    0.9818\\
17    0.9873\\
18    0.9921\\
19    0.9963\\
20    1.0000\\
21    1.0000\\
22    1.0000\\
23    1.0000\\
24    1.0000\\
};
\addlegendentry{$\bf{Algorithms\:1 \:\& \:2 \:at \: \textit{P} = 0\:\text{dB}}$}

\addplot [color=red, line width=1.0pt, mark=o, mark options={solid, red}]
  table[row sep=crcr]{%
0    0\\
1    0.3530\\
2    0.5132\\
3    0.6147\\
4    0.6866\\
5    0.7407\\
6    0.7832\\
7    0.8175\\
8    0.8458\\
9    0.8697\\
10    0.8901\\
11    0.9077\\
12   0.9231\\
13    0.9367\\
14    0.9487\\
15    0.9595\\
16    0.9692\\
17    0.9780\\
18    0.9860\\
19    0.9933\\
20    1.0000\\
21    1.0000\\
22    1.0000\\
23    1.0000\\
24    1.0000\\
};
\addlegendentry{$\bf{Algorithms\:1 \: \& \:2 \:at \: \textit{P} = 30\:\text{dB}}$}

\addplot [color=aqua, line width=1.0pt, mark=o, mark options={solid, aqua}]
  table[row sep=crcr]{%
0    0\\
1    0.4740\\
2    0.6360\\
3    0.7183\\
4    0.7684\\
5    0.8022\\
6    0.8267\\
7    0.8452\\
8    0.8599\\
9    0.8717\\
10    0.8815\\
11    0.8898\\
12    0.8969\\
13    0.9030\\
14    0.9084\\
15    0.9132\\
16    0.9174\\
17    0.9212\\
18    0.9247\\
19    0.9278\\
20    0.9307\\
21    0.9333\\
22    0.9358\\
23    0.9380\\
24    0.9401\\
25    0.9421\\
26    0.9439\\
27    0.9456\\
28    0.9472\\
29    0.9487\\
30    0.9501\\
31    0.9514\\
32    0.9527\\
33    0.9539\\
34    0.9550\\
35    0.9561\\
36    0.9572\\
37    0.9581\\
38    0.9591\\
39    0.9600\\
40    0.9609\\
41    0.9617\\
42    0.9625\\
43    0.9632\\
44    0.9640\\
45    0.9647\\
46    0.9653\\
47    0.9660\\
48    0.9666\\
49    0.9672\\
50    0.9678\\
51    0.9684\\
52    0.9689\\
53    0.9695\\
54    0.9700\\
55    0.9705\\
56    0.9710\\
57    0.9714\\
58    0.9719\\
59    0.9723\\
60    0.9728\\
61    0.9732\\
62    0.9736\\
63    0.9740\\
64    0.9744\\
65    0.9747\\
66    0.9751\\
67    0.9755\\
68    0.9758\\
69    0.9761\\
70    0.9765\\
71    0.9768\\
72    0.9771\\
73    0.9774\\
74    0.9777\\
75    0.9780\\
76    0.9783\\
77    0.9786\\
78    0.9788\\
79    0.9791\\
80    0.9794\\
81    0.9796\\
82    0.9799\\
83    0.9801\\
84    0.9804\\
85    0.9806\\
86    0.9808\\
87    0.9810\\
88    0.9813\\
89    0.9815\\
90    0.9817\\
91    0.9819\\
92    0.9821\\
93    0.9823\\
94    0.9825\\
95    0.9827\\
96    0.9829\\
97    0.9831\\
98    0.9833\\
99    0.9834\\
100    0.9836\\
};
\addlegendentry{$\bf{TTS\:(SSCA) \:at \: \textit{P} = 0\:\text{dB}}$}

\addplot [color=green, line width=1.0pt, mark=o, mark options={solid, green}]
  table[row sep=crcr]{%
0    0\\
1    0.4358\\
2    0.5960\\
3    0.6813\\
4    0.7347\\
5    0.7716\\
6    0.7986\\
7    0.8192\\
8    0.8356\\
9    0.8489\\
10    0.8599\\
11    0.8693\\
12    0.8772\\
13    0.8841\\
14    0.8902\\
15    0.8955\\
16    0.9003\\
17    0.9045\\
18    0.9083\\
19    0.9118\\
20    0.9150\\
21    0.9179\\
22    0.9206\\
23    0.9230\\
24    0.9253\\
25    0.9275\\
26    0.9294\\
27    0.9313\\
28    0.9330\\
29    0.9346\\
30    0.9362\\
31    0.9376\\
32    0.9390\\
33    0.9403\\
34    0.9415\\
35    0.9427\\
36    0.9438\\
37    0.9448\\
38    0.9458\\
39    0.9468\\
40    0.9477\\
41    0.9486\\
42    0.9495\\
43    0.9503\\
44    0.9511\\
45    0.9518\\
46    0.9525\\
47    0.9532\\
48    0.9539\\
49    0.9545\\
50    0.9552\\
51    0.9558\\
52    0.9564\\
53    0.9569\\
54    0.9575\\
55    0.9580\\
56    0.9585\\
57    0.9590\\
58    0.9595\\
59    0.9600\\
60    0.9605\\
61    0.9609\\
62    0.9613\\
63    0.9618\\
64    0.9622\\
65    0.9626\\
66    0.9630\\
67    0.9634\\
68    0.9637\\
69    0.9641\\
70    0.9645\\
71    0.9648\\
72    0.9652\\
73    0.9655\\
74    0.9658\\
75    0.9662\\
76    0.9665\\
77    0.9668\\
78    0.9671\\
79    0.9674\\
80    0.9677\\
81    0.9680\\
82    0.9682\\
83    0.9685\\
84    0.9688\\
85    0.9690\\
86    0.9693\\
87    0.9696\\
88    0.9698\\
89    0.9701\\
90    0.9703\\
91    0.9705\\
92    0.9708\\
93    0.9710\\
94    0.9712\\
95    0.9714\\
96    0.9717\\
97    0.9719\\
98    0.9721\\
99    0.9723\\
100    0.9725\\
};
\addlegendentry{$\bf{TTS\:(SSCA) \:at \: \textit{P} = 30\:\text{dB}}$}

\end{axis}

% \begin{axis}[%
% width=1.227\figW,
% height=1.227\figH,
% at={(-0.16\figW,-0.135\figH)},
% scale only axis,
% xmin=0,
% xmax=1,
% ymin=0,
% ymax=1,
% axis line style={draw=none},
% ticks=none,
% axis x line*=bottom,
% axis y line*=left
% ]
% \end{axis}
\end{tikzpicture}%
} 
		\subcaption{}
		\label{convergence_K1_compare}
  \end{minipage}
  \caption{Convergence Plot of the Proposed Algorithms (a) and with the Normalised Objective~(b) for $K$~=~$1$}
\end{figure}
\subsection{Single-User Case}
We first consider the single-user case where the user is placed at ($D\,\text{m}, 0\,\text{m}$). In order to compare the proposed algorithms for the single-user case with the single-user algorithms in \cite{syed2023design} exploiting the statistical channel knowledge, we consider $\bfT' = \bf{0}$ similar to \cite{syed2023design}. At first, the convergence plot of the long-term optimisation part of the proposed algorithms is presented in Fig.~\ref{convergence_K1} for a single scenario of covariance matrices with $D = 30\,\text{m}$ and the transmit power levels of $P = 0$ dB and $P~=~30$ dB. Both Algorithm~1 and Algorithm~2 have the same long-term optimisation method. The convergence analysis reveals that the proposed algorithms converge in approximately 20 iterations for both lower and higher power levels, i.e., the convergence behaviour is independent of the transmit power level. We also compare the convergence of our algorithms to the SSCA based TTS method of~\cite{twotime} in Fig.~\ref{convergence_K1_compare}. Since the algorithms have different metrics for optimisation, we normalise the objective values between 0 and 1, where 0 denotes the minimum objective value and 1 denotes the maximum value over the iterations. It is clear from Fig.~\ref{convergence_K1_compare} that the proposed algorithms have a faster rate of convergence than the TTS method of~\cite{twotime}, which takes more than 50 iterations to converge.   
\par
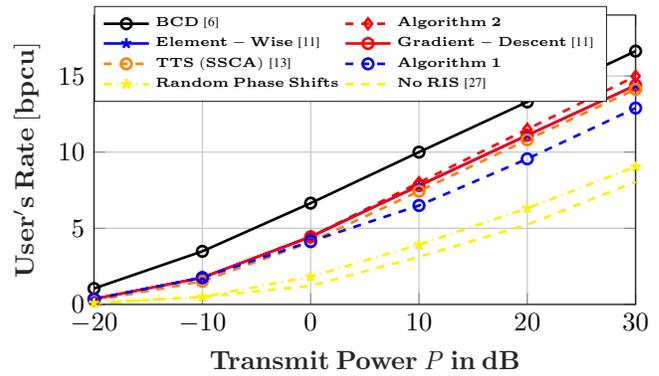
\begin{figure}		
\definecolor{mycolor1}{rgb}{0.00000,0.44700,0.74100}%
\definecolor{mycolor2}{rgb}{0.85000,0.32500,0.09800}%
\definecolor{mycolor3}{rgb}{0.92900,0.69400,0.12500}%
\definecolor{mycolor4}{rgb}{0.49400,0.18400,0.55600}%
\definecolor{mycolor5}{rgb}{0.46600,0.67400,0.18800}%
\definecolor{mycolor6}{rgb}{0.30100,0.74500,0.93300}%
\definecolor{mycolor7}{rgb}{0.63500,0.07800,0.18400}%
\begin{tikzpicture}

\begin{axis}[%
width=0.9\figW,
height=\figH,
at={(0\figW,0\figH)},
scale only axis,
xmin=-20,
xmax=30,
xlabel style={font=\color{white!15!black}},
xlabel={$\bf{Transmit\:Power\:{\it{P}}\:in\:dB}$},
ymin=0,
ymax=19,
ylabel style={font=\color{white!15!black}},
ylabel={$\bf{User's\:Rate\:[bpcu]}$},
axis background/.style={fill=white},
xmajorgrids,
ymajorgrids,
legend style={at={(0,0.70)}, anchor=south west, legend cell align=left, align=left, draw=white!15!black, row sep=-0.05cm, font=\tiny},
legend columns=2
]
\addplot [color=black, solid, line width=1.0pt, mark=o, mark options={solid, black}]
  table[row sep=crcr]{%
-20	1.04576250847525\\
-10	3.48694842975115\\
0	6.65022601517131\\
10	9.99992412671442\\
20	13.2966182468131\\
30	16.6318245174515\\
}; 
\addlegendentry{$\bf{BCD}$ \cite{guo2020weighted}}

\addplot [color=red, dashed, line width=1.0pt, mark=diamond, mark options={solid, red}]
  table[row sep=crcr]{%
-20	0.354630718825827\\
-10	1.7671032366853\\
0	4.43836491270356\\
10	7.99642539263533\\
20	11.5064509863262\\
30	14.9847163747852\\
}; 
\addlegendentry{$\bf{Algorithm\:2}$}

\addplot [color=blue, line width=1.0pt, mark=star, mark options={solid, blue}]
  table[row sep=crcr]{%
-20	0.355876940226198\\
-10	1.76525565481984\\
0	4.44135503821663\\
10	7.79815325404131\\
20	11.1067470810331\\
30	14.3871707411701\\
};
\addlegendentry{$\bf{Element-Wise}$ \cite{syed2023design}}

\addplot [color=red, line width=1.0pt, mark=o, mark options={solid, red}]
  table[row sep=crcr]{%
-20	0.3560127540565\\
-10	1.76497886220261\\
0	4.44157005393869\\
10	7.79839437714246\\
20	11.1067556718863\\
30	14.3871167582596\\
};
\addlegendentry{$\bf{Gradient-Descent}$  \cite{syed2023design}}

\addplot [color=orange, dashed, line width=1.0pt, mark=o, mark options={solid, orange}]
  table[row sep=crcr]{%
-20	0.258916514265459\\
-10	1.52235460633307\\
0	4.09621315345868\\
10	7.43706740422075\\
20	10.8096892229566\\
30	14.1227755562493\\
};
\addlegendentry{$\bf{TTS~(SSCA)}$ \cite{twotime}}

\addplot [color=blue, dashed, line width=1.0pt, mark=o, mark options={solid, blue}]
  table[row sep=crcr]{%
-20	0.354566299480666\\
-10	1.75320948198134\\
0	4.13770103964169\\
10	6.49607317816577\\
20	9.55959544311597\\
30	12.8927965755973\\
};
\addlegendentry{$\bf{Algorithm\:1}$}

%\addplot [color=green, dashdotted, line width=1.0pt, mark=o, mark options={solid, green}]
%  table[row sep=crcr]{%
%-20	0.0978485540359919\\
%-10	0.586082542331212\\
%0	2.2213950008991\\
%10	4.97921680113047\\
%20	8.24438671218271\\
%30	11.7443831113453\\
%};
%\addlegendentry{$\bf{AO\:Algorithm\:}$ \cite{dang2020joint}}

\addplot [color=yellow, dashdotted, line width=1.0pt, mark=star, mark options={solid, yellow}]
  table[row sep=crcr]  {%
-20	0.0870485540355525\\
-10	0.496082542331212\\
0	1.8213950008991\\
10	3.92101680111174\\
20	6.3124438671218271\\
30	9.0443831113453\\
};
\addlegendentry{$\bf{Random\:Phase\:Shifts}$}

\addplot [color=yellow, dashed, line width=1.0pt]
  table[row sep=crcr]{%
-20	0.0870485540355525\\
-10	0.496082542331212\\
0	1.2213950008991\\
10	3.12101680111174\\
20	5.24438671218271\\
30	8.0443831113453\\
};
\addlegendentry{$\bf{No\:RIS}$ \cite{amor2020bilinear}}

\end{axis}

% \begin{axis}[%
% width=1.227\figW,
% height=1.227\figH,
% at={(-0.16\figW,-0.135\figH)},
% scale only axis,
% xmin=0,
% xmax=1,
% ymin=0,
% ymax=1,
% axis line style={draw=none},
% ticks=none,
% axis x line*=bottom,
% axis y line*=left
% ]
% \end{axis}
\end{tikzpicture}% 
		\caption{User's Rate vs Transmit Power $P$ in dB for $K=1$}
		\label{rate 1 user}
\end{figure}

 The user's rate is next plotted in Fig.~\ref{rate 1 user}, which is computed with the different algorithms and compared over the transmit power levels $P$ with $D$ varying in between $15\,\text{m}$ to $60\,\text{m}$. The performance of the proposed algorithms is compared with the algorithms in \cite{syed2023design, twotime} exploiting the statistical channel knowledge, and also with \cite{guo2020weighted} employing the instantaneous CSI for both the filter and the phase shift optimisation. Fig.~\ref{rate 1 user} shows that the algorithm in \cite{guo2020weighted} serves as an upper bound to the algorithms using only the statistical information for the phase shift design. Both of the proposed algorithms have a similar performance at low $P$, however, Algorithm~1, employing the bilinear precoder, is not optimal at the higher power levels~\cite{amor2020bilinear}. Hence, Algorithm~2 with the optimal filters computed using the BCD method outperforms Algorithm~1 at high $P$ and at the price of higher complexity. The proposed algorithms clearly offer a considerable performance gain when compared to a system without RIS \cite{amor2020bilinear} or to the one employing RIS with random phase shifts. Fig.~\ref{rate 1 user} also shows that Algorithm~2 slightly outperforms the TTS approach in \cite{twotime} based on the SSCA method, which also uses the optimal precoding filters computed via the IWMMSE method \cite{wmmse} in every channel coherence interval. Moreover, %these algorithms based on maximising the lower bound of the user's rate~\cite{medard} outperform the AO algorithm in~\cite{dang2020joint}, which maximises the upper bound of the rate obtained through Jensen's inequality. Furthermore, 
Algorithm~2 employing the hybrid design with the optimal BCD filters also slightly outperforms the hybrid algorithms with the optimal filters in~\cite{syed2023design} at high transmit power levels. 
\subsection{Multi-User Case}
We next consider a multi-user system with $K = 3$ users, randomly placed inside a circle of radius $50\,\text{m}$ at a distance $D = 30\,\text{m}$ from the BS. At first, the convergence plot of the proposed algorithms for the multi-user case is shown in  Fig.~\ref{convergence_K3}. Similar to the single-user case, the convergence of the algorithms is independent of the transmit power level and they converge in approximately 25 iterations. We again compare the rate of convergence of our algorithms to the SSCA based TTS method of~\cite{twotime} in Fig.~\ref{convergence_norm}. The objective values of the algorithms are normalised between 0 and 1 for comparison. The proposed algorithms have a faster rate of convergence than the TTS method of~\cite{twotime}, which takes more than 100 iterations to converge, and the difference in rate is even larger here than the single-user case in Fig.~\ref{convergence_K1_compare}. 
\begin{figure}%[ht]	
\begin{minipage}{.25\textwidth}
		 \scalebox{0.5}{% This file was created by matlab2tikz.
\definecolor{mycolor1}{rgb}{0.00000,0.44700,0.74100}%
\definecolor{mycolor2}{rgb}{0.85000,0.32500,0.09800}%
\begin{tikzpicture}

\begin{axis}[%
width=0.85\figW,
height=\figH,
at={(0\figW,0\figH)},
scale only axis,
xmin=0,
xmax=30,
xlabel style={font=\color{white!15!black}},
xlabel={$\bf{Iteration\:No.}$},
ymin=0.1,
ymax=7,
ylabel style={font=\color{white!15!black}},
ylabel={$\bf{Lower\:Bound\:of\:Rate\:[bpcu]}$},
axis background/.style={fill=white},
xmajorgrids,
ymajorgrids,
legend style={at={(0.95,0.2)}, anchor=south east, legend cell align=left, align=left, draw=white!15!black, row sep=-0.05cm}
]
\addplot [color=blue, line width=1.0pt, mark=diamond, mark options={solid, blue}]
  table[row sep=crcr]{%
0	2.40498708302359\\
1	3.88687769005029\\
2	4.36958808650592\\
3	4.63959529393905\\
4	4.81441974975845\\
5	4.9370538178702\\
6	5.02763917491674\\
7	5.09702416546028\\
8	5.15161833097356\\
9	5.19547065301731\\
10	5.23127285528822\\
11	5.2608881836621\\
12	5.28564953877601\\
13	5.30653686065268\\
14	5.32428741981823\\
15	5.33946696929754\\
16	5.35251711137701\\
17	5.36378769495486\\
18	5.37355950548181\\
19	5.38206049219489\\
20	5.38947759110851\\
21	5.38947759110851\\
22	5.38947759110851\\
23	5.38947759110851\\
24	5.38947759110851\\
};
\addlegendentry{${P = 0\:\text{dB}}$}

\addplot [color=red, line width=1.0pt, mark=o, mark options={solid, red}]
  table[row sep=crcr]{%
0	2.84803304613519\\
1	4.48500458512907\\
2	5.08651155047514\\
3	5.43517145775041\\
4	5.66825171847585\\
5	5.83599902013014\\
6	5.96294242677466\\
7	6.0625360874214\\
8	6.1428385346743\\
9	6.20899452877924\\
10	6.26445298659824\\
11	6.31161732255062\\
12	6.35221613777225\\
13	6.38752587644141\\
14	6.41851036152504\\
15	6.44591142524686\\
16	6.4703096150397\\
17	6.49216596571168\\
18	6.5118513711561\\
19	6.52966763580555\\
20	6.5458629479322\\
21	6.5458629479322\\
22	6.5458629479322\\
23	6.5458629479322\\
24	6.5458629479322\\
};
\addlegendentry{${P = 30\:\text{dB}}$}

\end{axis}

% \begin{axis}[%
% width=1.227\figW,
% height=1.227\figH,
% at={(-0.16\figW,-0.135\figH)},
% scale only axis,
% xmin=0,
% xmax=1,
% ymin=0,
% ymax=1,
% axis line style={draw=none},
% ticks=none,
% axis x line*=bottom,
% axis y line*=left
% ]
% \end{axis}
\end{tikzpicture}%
}
		\subcaption{}
		\label{convergence_K3}
  \end{minipage}%
 \begin{minipage}{.3\textwidth}
	\scalebox{0.5}{% This file was created by matlab2tikz.
\definecolor{mycolor1}{rgb}{0.00000,0.44700,0.74100}%
\definecolor{mycolor2}{rgb}{0.85000,0.32500,0.09800}%
\definecolor{aqua}{rgb}{0.0, 1.0, 1.0}
\begin{tikzpicture}

\begin{axis}[%
width=0.85\figW,
height=\figH,
at={(0\figW,0\figH)},
scale only axis,
xmin=0,
xmax=100, %200
xlabel style={font=\color{white!15!black}},
xlabel={$\bf{Iteration\:No.}$},
ymin=0.1,
ymax=1,
ylabel style={font=\color{white!15!black}},
ylabel={$\bf{Normalised\:Objective}$},
axis background/.style={fill=white},
xmajorgrids,
ymajorgrids,
legend style={at={(0.95,0.01)}, anchor=south east, legend cell align=left, align=left, draw=white!15!black, row sep=-0.05cm}
]
\addplot [color=blue, line width=1.0pt, mark=diamond, mark options={solid, blue}]
  table[row sep=crcr]{%
0    0\\
1    0.4965\\
2   0.6583\\
3    0.7487\\
4    0.8073\\
5    0.8484\\
6    0.8788\\
7    0.9020\\
8    0.9203\\
9    0.9350\\
10    0.9470\\
11   0.9569\\
12    0.9652\\
13    0.9722\\
14    0.9782\\
15    0.9832\\
16   0.9876\\
17    0.9914\\
18    0.9947\\
19    0.9975\\
20    1.0000\\
21    1.0000\\
22   1.0000\\
23    1.0000\\
24    1.0000\\
};
\addlegendentry{$\bf{Algorithms\:1 \:\& \:2 \:at \: \textit{P} = 0\:\text{dB}}$}

\addplot [color=red, line width=1.0pt, mark=o, mark options={solid, red}]
  table[row sep=crcr]{%
0    0\\
1    0.4427\\
2    0.6053\\
3    0.6996\\
4    0.7627\\
5    0.8080\\
6    0.8424\\
7    0.8693\\
8    0.8910\\
9    0.9089\\
10    0.9239\\
11   0.9367\\
12    0.9476\\
13    0.9572\\
14    0.9656\\
15    0.9730\\
16    0.9796\\
17    0.9855\\
18    0.9908\\
19    0.9956\\
20    1.0000\\
21    1.0000\\
22    1.0000\\
23    1.0000\\
24    1.0000\\
};
\addlegendentry{$\bf{Algorithms\:1 \: \& \:2 \:at \: \textit{P} = 30\:\text{dB}}$}

\addplot [color=aqua, line width=1.0pt, mark=o, mark options={solid, aqua}]
  table[row sep=crcr]{%
0    0\\
1    0.4203\\
2    0.5675\\
3    0.6434\\
4    0.6899\\
5    0.7214\\
6    0.7444\\
7    0.7626\\
8    0.7770\\
9    0.7886\\
10    0.7983\\
11   0.8065\\
12    0.8135\\
13    0.8196\\
14    0.8250\\
15    0.8298\\
16    0.8343\\
17    0.8383\\
18    0.8420\\
19    0.8454\\
20    0.8485\\
21    0.8515\\
22    0.8542\\
23    0.8566\\
24    0.8580\\
25    0.8593\\
26    0.8614\\
27    0.8634\\
28    0.8653\\
29    0.8672\\
30    0.8690\\
31    0.8707\\
32    0.8724\\
33    0.8740\\
34    0.8755\\
35    0.8770\\
36    0.8785\\
37    0.8799\\
38    0.8813\\
39    0.8826\\
40    0.8840\\
41    0.8852\\
42    0.8865\\
43    0.8877\\
44    0.8888\\
45    0.8900\\
46    0.8911\\
47    0.8922\\
48    0.8933\\
49    0.8943\\
50    0.8953\\
51    0.8963\\
52    0.8973\\
53    0.8982\\
54    0.8992\\
55    0.9001\\
56    0.9010\\
57    0.9018\\
58    0.9027\\
59    0.9032\\
60    0.9041\\
61    0.9049\\
62    0.9057\\
63    0.9065\\
64    0.9073\\
65    0.9080\\
66    0.9088\\
67    0.9095\\
68    0.9103\\
69    0.9110\\
70    0.9117\\
71    0.9124\\
72    0.9131\\
73    0.9138\\
74    0.9145\\
75    0.9151\\
76    0.9158\\
77    0.9165\\
78    0.9172\\
79    0.9179\\
80    0.9186\\
81    0.9192\\
82    0.9199\\
83    0.9205\\
84    0.9212\\
85    0.9218\\
86    0.9224\\
87    0.9230\\
88    0.9236\\
89    0.9242\\
90    0.9248\\
91    0.9254\\
92    0.9260\\
93    0.9266\\
94    0.9271\\
95    0.9277\\
96    0.9282\\
97    0.9288\\
98    0.9293\\
99    0.9298\\
100    0.9304\\
101    0.9309\\
102    0.9314\\
103    0.9319\\
104    0.9324\\
105    0.9329\\
106    0.9334\\
107    0.9339\\
108    0.9344\\
109    0.9348\\
110   0.9353\\
111    0.9358\\
112    0.9362\\
113    0.9367\\
114    0.9371\\
115    0.9376\\
116    0.9380\\
117    0.9385\\
118    0.9389\\
119    0.9393\\
120    0.9397\\
121    0.9402\\
122    0.9406\\
123    0.9410\\
124    0.9414\\
125    0.9418\\
126    0.9422\\
127    0.9426\\
128    0.9430\\
129    0.9434\\
130    0.9438\\
131    0.9442\\
132    0.9445\\
133    0.9449\\
134    0.9453\\
135    0.9457\\
136    0.9460\\
137    0.9464\\
138    0.9468\\
139    0.9471\\
140    0.9475\\
141    0.9478\\
142    0.9482\\
143    0.9485\\
144    0.9488\\
145    0.9492\\
146    0.9495\\
147    0.9499\\
148    0.9502\\
149    0.9505\\
150    0.9508\\
151    0.9512\\
152    0.9515\\
153    0.9518\\
154    0.9521\\
155    0.9524\\
156    0.9527\\
157    0.9530\\
158    0.9533\\
159    0.9536\\
160    0.9539\\
161    0.9542\\
162    0.9545\\
163    0.9548\\
164    0.9551\\
165    0.9554\\
166    0.9557\\
167    0.9560\\
168    0.9563\\
169    0.9565\\
170    0.9568\\
171    0.9571\\
172    0.9574\\
173    0.9576\\
174    0.9579\\
175    0.9582\\
176    0.9584\\
177    0.9587\\
178    0.9590\\
179    0.9592\\
180    0.9595\\
181    0.9597\\
182    0.9600\\
183    0.9603\\
184    0.9605\\
185    0.9608\\
186    0.9610\\
187    0.9613\\
188    0.9615\\
189    0.9617\\
190    0.9620\\
191    0.9622\\
192    0.9625\\
193    0.9627\\
194    0.9629\\
195    0.9632\\
196    0.9634\\
197    0.9636\\
198    0.9639\\
199    0.9641\\
};
\addlegendentry{$\bf{TTS\:(SSCA) \:at \: \textit{P} = 0\:\text{dB}}$}

\addplot [color=green, line width=1.0pt, mark=o, mark options={solid, green}]
  table[row sep=crcr]{%
0     0\\
1    0.2581\\
2    0.3859\\
3    0.4631\\
4    0.5176\\
5    0.5594\\
6    0.5929\\
7    0.6204\\
8    0.6431\\
9    0.6623\\
10    0.6787\\
11   0.6929\\
12    0.7053\\
13    0.7163\\
14    0.7261\\
15    0.7350\\
16    0.7430\\
17    0.7505\\
18    0.7574\\
19    0.7638\\
20    0.7698\\
21    0.7755\\
22    0.7810\\
23    0.7861\\
24    0.7910\\
25    0.7957\\
26    0.8003\\
27    0.8046\\
28    0.8088\\
29    0.8128\\
30    0.8167\\
31    0.8205\\
32    0.8241\\
33    0.8277\\
34    0.8311\\
35    0.8344\\
36    0.8375\\
37    0.8406\\
38    0.8436\\
39    0.8465\\
40    0.8493\\
41    0.8521\\
42    0.8547\\
43    0.8573\\
44    0.8598\\
45    0.8622\\
46    0.8645\\
47    0.8668\\
48    0.8690\\
49    0.8711\\
50    0.8732\\
51    0.8753\\
52    0.8772\\
53    0.8791\\
54    0.8810\\
55    0.8828\\
56    0.8846\\
57    0.8863\\
58    0.8880\\
59    0.8896\\
60    0.8912\\
61    0.8928\\
62    0.8943\\
63    0.8958\\
64    0.8972\\
65    0.8986\\
66    0.9000\\
67    0.9014\\
68    0.9027\\
69    0.9040\\
70    0.9052\\
71    0.9065\\
72    0.9077\\
73    0.9089\\
74    0.9100\\
75    0.9111\\
76    0.9122\\
77    0.9133\\
78    0.9144\\
79    0.9154\\
80    0.9164\\
81    0.9174\\
82    0.9184\\
83    0.9194\\
84    0.9203\\
85    0.9212\\
86    0.9222\\
87    0.9230\\
88    0.9239\\
89    0.9248\\
90    0.9256\\
91    0.9264\\
92    0.9273\\
93    0.9280\\
94    0.9288\\
95    0.9296\\
96    0.9304\\
97    0.9311\\
98    0.9318\\
99    0.9326\\
100    0.9333\\
101    0.9340\\
102    0.9346\\
103    0.9353\\
104    0.9360\\
105    0.9366\\
106    0.9373\\
107    0.9379\\
108    0.9385\\
109    0.9391\\
110    0.9397\\
111    0.9403\\
112    0.9409\\
113    0.9415\\
114    0.9420\\
115    0.9426\\
116    0.9431\\
117    0.9437\\
118    0.9442\\
119    0.9448\\
120    0.9453\\
121    0.9458\\
122    0.9463\\
123    0.9468\\
124    0.9473\\
125    0.9478\\
126    0.9482\\
127    0.9487\\
128    0.9492\\
129    0.9496\\
130    0.9501\\
131    0.9505\\
132    0.9510\\
133    0.9514\\
134    0.9518\\
135    0.9522\\
136    0.9527\\
137    0.9531\\
138    0.9535\\
139    0.9539\\
140    0.9543\\
141    0.9547\\
142    0.9550\\
143    0.9554\\
144    0.9558\\
145    0.9562\\
146    0.9565\\
147    0.9569\\
148    0.9573\\
149    0.9576\\
150    0.9580\\
151    0.9583\\
152    0.9587\\
153    0.9590\\
154    0.9593\\
155    0.9597\\
156    0.9600\\
157    0.9603\\
158    0.9606\\
159    0.9610\\
160    0.9613\\
161    0.9616\\
162    0.9619\\
163    0.9622\\
164    0.9625\\
164    0.9628\\
165    0.9631\\
166    0.9634\\
167    0.9637\\
168    0.9639\\
169    0.9642\\
170    0.9645\\
171    0.9648\\
172    0.9651\\
173    0.9653\\
174    0.9656\\
175    0.9659\\
176    0.9661\\
177    0.9664\\
178    0.9666\\
179    0.9669\\
180    0.9671\\
181    0.9674\\
182    0.9676\\
183    0.9679\\
184    0.9681\\
185    0.9684\\
186    0.9686\\
187    0.9688\\
188    0.9691\\
189    0.9693\\
190    0.9695\\
191    0.9698\\
192    0.9700\\
193    0.9702\\
194    0.9704\\
195    0.9706\\
196    0.9709\\
197    0.9711\\
198    0.9713\\
};
\addlegendentry{$\bf{TTS\:(SSCA) \:at \: \textit{P} = 30\:\text{dB}}$}

\end{axis}

% \begin{axis}[%
% width=1.227\figW,
% height=1.227\figH,
% at={(-0.16\figW,-0.135\figH)},
% scale only axis,
% xmin=0,
% xmax=1,
% ymin=0,
% ymax=1,
% axis line style={draw=none},
% ticks=none,
% axis x line*=bottom,
% axis y line*=left
% ]
% \end{axis}
\end{tikzpicture}%
 }
		\subcaption{}
		\label{convergence_norm}
  \end{minipage}
  \caption{Convergence Plot of the Proposed Algorithms (a) and with the Normalised Objective~(b) for $K$~=~$3$}
\end{figure}
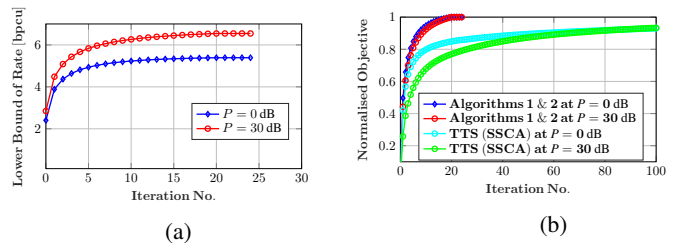
 \begin{figure}%[ht]
\definecolor{mycolor1}{rgb}{1.00000,0.00000,1.00000}%
\definecolor{mycolor2}{rgb}{0.00000,1.00000,1.00000}%
\begin{tikzpicture}

\begin{axis}[%
width=0.9\figW,
height=\figH,
at={(0\figW,0\figH)},
scale only axis,
xmin=-20,
xmax=30,
xlabel style={font=\color{white!15!black}},
xlabel={$\bf{Transmit\:Power\:in\:dB}$},
ymin=0,
ymax=35,
ylabel style={font=\color{white!15!black}},
ylabel={$\bf{Sum-Rate\:[bpcu]}$},
axis background/.style={fill=white},
xmajorgrids,
ymajorgrids,
legend style={at={(0, 1)}, anchor=north west, legend cell align=left, align=left, draw=white!15!black, row sep=-0.05cm, font=\tiny}
]

\addplot [color=mycolor1, line width=1.0pt, mark=triangle, mark options={solid, rotate=90, mycolor1}]
  table[row sep=crcr]{%
-20	1.59933865325028\\
-10	4.30223236830499\\
0	7.7948815704052\\
10	12.2482506856857\\
20	20.082854833368\\
30	29.2765221709976\\
};
\addlegendentry{$\bf{BCD}$ \cite{guo2020weighted}}

\addplot [color=red, solid, line width=1.0pt, mark=diamond, mark options={solid, red}]
  table[row sep=crcr]{%
-20	0.75781393951688\\
-10	2.66125956093942\\
0	4.43414698972493\\
10	10.2220245299256\\
20	17.9975328337698\\
30	26.9517708731908\\
};
\addlegendentry{$\bf{Algorithm\:2\:(BCD)}$}

\addplot [color=mycolor2, line width=1.0pt, mark=asterisk, mark options={solid, mycolor2}]
  table[row sep=crcr]{%
-20	0.125710800721553\\
-10	0.751658661509831\\
0	2.80111703431985\\
10	9.55656061400021\\
20	17.8218261676822\\
30	27.1085922454674\\
};
\addlegendentry{$\bf{Algorithm\:2\:(ZF)}$}

\addplot [color=green, line width=1.0pt, mark=asterisk, mark options={solid, green}]
  table[row sep=crcr]{%
-20	1.23772699873812\\
-10	1.70062016370718\\
0	4.96522868419965\\
10	8.7829072390171\\
20	16.5621307956317\\
30	25.7149387056552\\
};
\addlegendentry{$\bf{TTS~(SSCA)}$ \cite{twotime}}

\addplot [color=blue, solid, line width=1.0pt, mark=o, mark options={solid, blue}]
  table[row sep=crcr] {%
-20	0.665093988703588\\
-10	1.90969862619835\\
0	3.45751235932233\\
10	6.70356926896041\\
20	7.90913455649914\\
30	8.7190888779241\\
};
\addlegendentry{$\bf{Algorithm\:1}$}

\addplot [color=black, dashdotted, line width=1.0pt, mark=triangle, mark options={solid, rotate=90, black}]
  table[row sep=crcr]{%
-20	0.125332370141192\\
-10	0.874739207070896\\
0	3.26653413966587\\
10	6.87161381640299\\
20	14.0502834348178\\
30	21.6158683379878\\
};
\addlegendentry{$\bf{Random\:Phase\:(BCD)}$}

\addplot [color=yellow, dashdotted, line width=1.0pt, mark=triangle, mark options={solid, rotate = 90, yellow}]
  table[row sep=crcr]{%
-20	0.0389321604441677\\
-10	0.337902616335319\\
0	1.82694532195457\\
10	5.74140656528051\\
20	12.3665553206194\\
30	18.7636340960608\\
};
\addlegendentry{$\bf{No\:RIS\:(BCD)}$}

\addplot [color=black, dashed, line width=1.0pt, mark=asterisk, mark options={solid, black}]
  table[row sep=crcr]{%
-20	0.106022191745722\\
-10	0.411834654774041\\
0	0.80829300915252\\
10	1.68344498421662\\
20	2.67655742000319\\
30	3\\
};
\addlegendentry{$\bf{Random\:Phase\:(GMF)}$}
\addplot [color=yellow, dashed, line width=1.0pt, mark=asterisk, mark options={solid, yellow}]
  table[row sep=crcr]{%
-20	0.09106022191745722\\
-10	0.35411834654774041\\
0	0.7280829300915252\\
10	1.48344498421662\\
20	2.07655742000319\\
30	2.7\\
};
\addlegendentry{$\bf{No\:RIS\:(GMF)}$}

 % Marking the point at x=20 with a red oval shape
\draw[gray, thick] (axis cs:20, 17) ellipse [x radius=0.5, y radius=2.6];

% Adding an arrow
\draw[->,ultra thick, gray, fill=white] (axis cs:19.5,16) to [out=150,in=-100] (axis cs:18.5,22.5);

\end{axis}

% \begin{axis}[%
% width=1.227\figW,
% height=1.227\figH,
% at={(-0.16\figW,-0.135\figH)},
% scale only axis,
% xmin=0,
% xmax=1,
% ymin=0,
% ymax=1,
% axis line style={draw=none},
% ticks=none,
% axis x line*=bottom,
% axis y line*=left
% ]
% \end{axis}
\end{tikzpicture}%
    \llap{\raisebox{3.5cm}{\scalebox{0.4}{% This file was created by matlab2tikz.
\definecolor{mycolor1}{rgb}{0.00000,0.44700,0.74100}%
\begin{tikzpicture}

\begin{axis}[%
width=0.9\figW,
height=\figH,
at={(0\figW,0\figH)},
scale only axis,
xmin=1,
xmax=100,
xtick={1, 10, 30, 50, 100},
x label style={at={(axis description cs:0.3,-0.1)},anchor=north},
xlabel={$\bf{Run\:-\:Time\:Ratio}$ },
ymin=0,
ymax=25, %32
y label style={at={(axis description cs:-0.1,0.28)},anchor=south, font=\color{white!15!black}},
ylabel={$\bf{\#\:Occurrences}$},
axis background/.style={fill=white},
%xmajorgrids,
ymajorgrids,
legend style={at={(0, 1)}, anchor=north west, legend cell align=left, align=left, draw=white!15!black, row sep=-0.05cm, font=\tiny}
%legend cell align={left},
%legend style={fill opacity=1, draw opacity=1, text opacity=1, at={(0.03,2)}, anchor=north west},
%tick align=inside,
%tick pos=left,
]
\addplot[ybar interval, fill=mycolor1, fill opacity=0.6, draw=black, area legend] table[row sep=crcr] {%
x	y\\
10	21\\
20	18\\
30	5\\
40	6\\
50	3\\
60	2\\
70	1\\
80	0\\
90	2\\
100	 0\\
110	1\\
120	0\\
130	0\\
140	0\\
150	0\\
160	0\\
170	0\\
180	0\\
190	0\\
200	0\\
210   1\\
};
\end{axis}

\end{tikzpicture}%
}}}
   \caption{Sum-Rate vs Transmit Power $P$ in dB for $K=3$}
   \label{plot2}
   \end{figure}
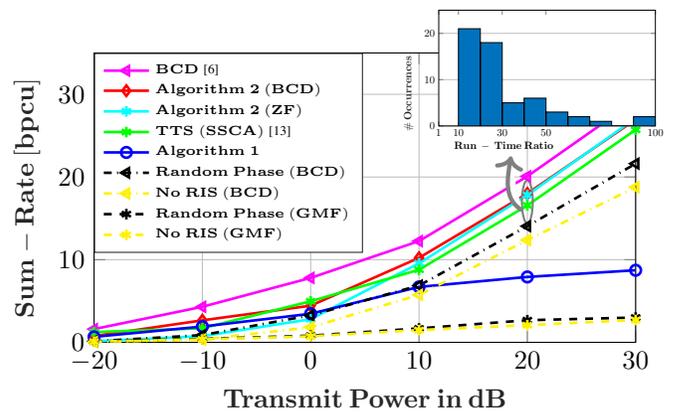
%    \begin{figure}%[ht]
% \begin{minipage}{0.8\textwidth}%[ht]
%    \scalebox{0.4}{\input{histogram}}
%    \caption{Histogram of the Run-Time Ratios of TTS~(SSCA) to Algorithms 1 $\&$ 2 at $P$ = 20 dB}
%    \label{hist}
% \end{minipage}
% \end{figure}
 \par The sum-rate of the users is plotted in Fig.~\ref{plot2}. The algorithm in \cite{guo2020weighted} performing the optimisation of both the filters and the phase shifts in every channel coherence interval acts as the upper bound to the algorithms involving only the statistical channel knowledge for the optimisation of the phase shifts. It can be seen that Algorithm~1 using the bilinear precoders offers a good performance only for low $P$ and the sum-rate starts to saturate gradually when the transmit power is higher than $10$~dB. This sum-rate saturation in the multi-user scenario is owing to the fact that the GMF cannot mitigate the inter-user interference. It can also be seen that Algorithm~2 is able to solve this saturation problem as it computes the optimal precoding filters in every channel coherence interval with the BCD method or the zero-forcing solution~(optimal at high $P$). It can be observed in Fig.~\ref{plot2} that the hybrid algorithm using the zero-forcing filters performs almost similar to the algorithm using the BCD method at high power levels, but there is some performance gap at lower power levels since the zero-forcing solution is not optimal when the noise level is high. Moreover, Algorithm~2 performs slightly better than the SSCA-based hybrid algorithm of \cite{twotime} on an average with the BCD filter and at high $P$ with the zero-forcing filter. All the algorithms are compared with systems having no RIS or having an RIS with random phase shifts. For these two baseline systems, the chosen filters are either the low-complexity bilinear precoders or those computed by the BCD method. Algorithm~2 with the optimal filters performs better than the baseline algorithms at all transmit power levels. The bilinear precoders based Algorithm~1 offers a performance gain when compared to the baseline algorithms employing the bilinear precoders \cite{amor2020bilinear}. However, it is only better than the baseline cases with the BCD filters at low $P$. This behaviour is expected and the difference in performance is attributed to the suboptimality of the low complexity bilinear precoders at high $P$ in comparison to the BCD filters. We also compare the simulation run-time of the long-term optimisation method of the proposed algorithms and SSCA-based TTS approach of \cite{twotime} for $P$ = $20$~dB in Fig.~\ref{plot2}. Since the run-time also depends on the processor, we only plot the ratio of run-time of \cite{twotime} to the proposed algorithms for comparison. From the histogram plot, it is observed that the ratio is always greater than 10, i.e. the proposed algorithms are an order of magnitude faster than the SSCA-based method even in the worst-case trials.     
%\vspace{-50pt}
 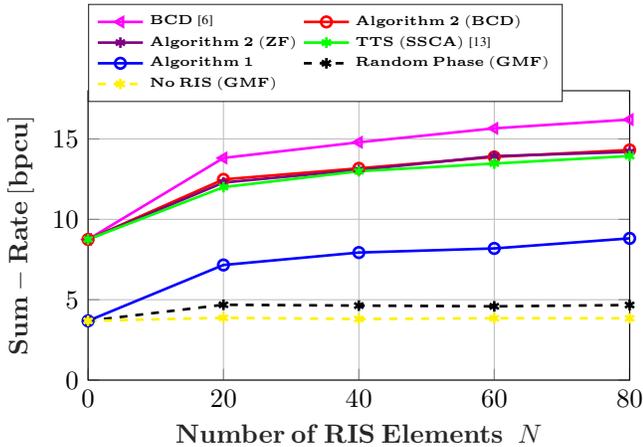
\begin{figure}%[ht]
\definecolor{mycolor1}{rgb}{1.00000,0.00000,1.00000}%
\definecolor{mycolor2}{rgb}{0.00000,1.00000,1.00000}%
\definecolor{mycolor3}{rgb}{0.92900,0.69400,0.12500}%
\definecolor{mycolor4}{rgb}{0.49400,0.18400,0.55600}%
\definecolor{mycolor5}{rgb}{0.46600,0.67400,0.18800}%
\definecolor{mycolor6}{rgb}{0.30100,0.74500,0.93300}%
\definecolor{mycolor7}{rgb}{0.63500,0.07800,0.18400}%
\begin{tikzpicture}

\begin{axis}[%
width=0.9\figW,
height=\figH,
at={(0\figW,0\figH)},
scale only axis,
xmin=0,
xmax=80,
xlabel style={font=\color{white!15!black}},
xlabel={$\bf{Number\:of\:RIS\:Elements\:}$ $N$},
ymin=0,
ymax=18, %32
ylabel style={font=\color{white!15!black}},
ylabel={$\bf{Sum-Rate\:[bpcu]}$},
axis background/.style={fill=white},
xmajorgrids,
ymajorgrids,
legend style={at={(0, 1.3)}, anchor=north west, legend cell align=left, align=left, draw=white!15!black, row sep=-0.05cm, font=\tiny}, legend columns = 2
%legend cell align={left},
%legend style={fill opacity=1, draw opacity=1, text opacity=1, at={(0.03,2)}, anchor=north west},
%tick align=inside,
%tick pos=left,
]
\addplot [color=mycolor1, line width=1.0pt, mark=triangle, mark options={solid, rotate=90, mycolor1}]
  table[row sep=crcr]{%
0    8.74919267057059 \\
20	13.8178342244386\\
40	14.7924734073837\\
60	15.6606636799726\\
80	16.2054475576151\\
};
\addlegendentry{$\bf{BCD}$ \cite{guo2020weighted}}

\addplot [color=red, line width=1.0pt, mark=o, mark options={solid, red}]
  table[row sep=crcr]{%
0    8.74919267057059 \\
20	12.4832936339083\\
40	13.1724754286821\\
60	13.8784252468825\\
80	14.3149925797329\\
};
\addlegendentry{$\bf{Algorithm\:2\:(BCD)}$}

\addplot [color=violet, line width=1.0pt, mark=asterisk, mark options={solid, violet}]
  table[row sep=crcr]{%
0    8.74919267057059 \\
20	12.2880837836174\\
40	13.0715988615438\\
60	13.9295589545188\\
80	14.1935712337321\\
};
\addlegendentry{$\bf{Algorithm\:2\:(ZF)}$}

\addplot [color=green, line width=1.0pt, mark=asterisk, mark options={solid, green}]
  table[row sep=crcr]{%
0    8.74919267057059 \\
20	12.0029376902037\\
40	12.9908078323959\\
60	13.4704665791515\\
80	13.9385506715734\\
};
\addlegendentry{$\bf{TTS~(SSCA)}$ \cite{twotime}}

\addplot [color=blue, solid, line width=1.0pt, mark=o, mark options={solid, blue}]
  table[row sep=crcr] {%
0	3.68524014444378\\
20	7.16010014755952\\
40	7.93395015913381\\
60	8.18990573632844\\
80	8.81606879069525\\
};
\addlegendentry{$\bf{Algorithm\:1}$}

%\addplot [color=mycolor6]
%  table[row sep=crcr]{%
%0    8.74919267057059 \\
%20	8.74919267057059\\
%40	8.74979565578218\\
%60	8.68167825919172\\
%80	8.81279482727179\\
%};
%\addlegendentry{No RIS BCD}

% \addplot [color=mycolor7]
%   table[row sep=crcr]{%
% 0	 8.74919267057059\\
% 20	9.6930646784623\\ %11 -> 9 
% 40	9.8460532247873\\
% 60	9.6667752736793\\
% 80	9.6667752736793\\
% };
% \addlegendentry{Random BCD}

\addplot [color=black, dashed, line width=1.0pt, mark=asterisk, mark options={solid, black}]
  table[row sep=crcr]{%
0	3.68524014444378\\
20	4.68524014444378\\
40	4.63379370959042\\
60	4.58758127388168\\
80	4.66877089107181\\
};
\addlegendentry{$\bf{Random\:Phase\:(GMF)}$}

\addplot [color=yellow, dashed, line width=1.0pt, mark=asterisk, mark options={solid, yellow}]
  table[row sep=crcr]{%
0	3.68524014444378\\
20	3.87481194244929\\
40	3.79974110901476\\
60	3.8491314380884\\
80	3.84132073396316\\
};
\addlegendentry{$\bf{No\:RIS\:(GMF)}$}

\end{axis}

% \begin{axis}[%
% width=5.833in,
% height=4.375in,
% at={(0in,0in)},
% scale only axis,
% xmin=0,
% xmax=1,
% ymin=0,
% ymax=1,
% axis line style={draw=none},
% ticks=none,
% axis x line*=bottom,
% axis y line*=left
% ]
% \end{axis}
% \begin{axis}[%
% width=1.227\figW,
% height=1.227\figH,
% at={(-0.16\figW,-0.135\figH)},
% scale only axis,
% xmin=0,
% xmax=1,
% ymin=0,
% ymax=1,
% axis line style={draw=none},
% ticks=none,
% axis x line*=bottom,
% axis y line*=left
% ]
% \end{axis}
\end{tikzpicture}%
    \caption{Sum-Rate vs $N$ at $P$ = 10 dB}
    \label{plot_N}
 \end{figure}
 \par Finally, Fig.~\ref{plot_N} shows the impact of the number of passive reflecting elements $N$ on the performance of the different algorithms at a fixed power $P = 10$ dB. It is observed that the sum-rate computed with the algorithm performing the optimisation of the phase shifts in every channel coherence interval \cite{guo2020weighted} grows at a faster rate with $N$ than the other lower complexity algorithms in Fig.~\ref{plot_N}. The performance of the proposed algorithms improves with an increase in $N$ at a similar rate. This behaviour is expected as both the algorithms employ the same optimisation procedure for the update of the phase shifts. Also, it is observed that the sum-rate does not improve with increasing $N$, if the phase shifts of the RIS are not optimised.  

\section{Conclusion}
This work has addressed the problem of the existing algorithms for an RIS-aided system relying on the knowledge of the instantaneous CSI for all the channels in every channel coherence interval. The information-theoretic lower bound of the users' sum-rate based on the worst-case noise is used as the figure of merit. This lower bound is then maximised to design the transmit filters at the BS and the phase shifts of the RIS. Our proposed algorithms, utilising the second-order channel statistics, significantly reduce the training overhead as the coherence interval of the covariance matrices consists of many channel coherence intervals. It is seen from the simulation results that there is a decline in the performance gain as compared to the algorithms optimising both the phase shifts and the transmit filters in every channel coherence interval, however, the proposed algorithms still bring a considerable improvement in the sum-rate of the users in comparison to the systems with no RIS or with random phase shifts. The first proposed algorithm uses the bilinear precoder and it performs all the optimisation processes only in the coherence interval of the channel statistics. However, it is observed to saturate in the high power regime for the multi-user case. The second algorithm employs the hybrid online/offline optimisation step and performs the transmit filter optimisation in every channel coherence interval, thereby having a higher complexity than the first algorithm. Nevertheless, it is seen to combat the observed saturation effect with the bilinear precoder. We have considered the covariance matrices to be perfectly known in this work. As a possible future work, one can focus on the effect of the imperfect covariance information on the performance of the RIS-aided communication systems. 

%\section{Appendix}
\appendix
\subsection{Lemma 1}
For any ${\bfu}\sim\nc({\bf{0}}, \bfC)$, it holds that
\begin{align*}
   \Ex\left[\bfu^{\Hm}\bfM_1\bfu \bfu^{\Hm} \bfM_2 \bfu \right] &= \tr(\bfC \bfM_1 \bfC \bfM_2) \\ &+ \tr(\bfC \bfM_1)\tr(\bfC \bfM_2). 
\end{align*}
\begin{proof}
 The lemma is a straightforward extension of Lemma~2 in \cite{bjornson2015massive}. 
%Let $\bfu = \bfC^{1/2}\Tilde{\bfu}$, where ${\Tilde{\bfu}}\sim\nc({\bf{0}}, {\bf{I}})$.
% \begin{align*}
%     \Ex\left[\bfu^{\Hm}\bfM_1\bfu \bfu^{\Hm} \bfM_2 \bfu \right] &= \Ex\Bigg[\Bigg(\sum\limits_{i,j}\Tilde{u}_i^{*}\left[\bfC^{\Hm/2}\bfM_1\bfC^{1/2}\right]_{i,j}\Tilde{u}_j \Bigg)\\
%     &\times\Bigg(\sum\limits_{k,l}\Tilde{u}_k^{*}\left[\bfC^{\Hm/2}\bfM_2\bfC^{1/2}\right]_{k,l}\Tilde{u}_l \Bigg) \Bigg]     
% \end{align*}
% Expanding the above terms and using the fact that $\Ex[\Tilde{u}_i^{*} \Tilde{u}_j] = 0$~$\forall \: i \neq j$, $\Ex[|\Tilde{u}_i|^2]~=~1$ and $\Ex[|\Tilde{u}_i|^4]~=~2$, we get the result in the lemma.
\end{proof}

\subsection{Proof of Theorem~1}
\begin{proof}
    \begin{align*}
    \mathrm{var}({\bfh_k^{\Hm}\bfp_k}) & = \mathop{{}\mathbb{E}}[|{{\bfh_k^{\Hm}}{\bfp}_k} -\mathop{{}\mathbb{E}}[\bfh_k^{\Hm}\bfp_k]|^2] \\
    &= \mathop{{}\mathbb{E}}\big[{\bfh_k^{\Hm}}\bfA_k{\bfh_k}{\bfh_k^{\Hm}}\bfA_k^{\Hm}\bfh_k\big] - \big|\mathrm{tr}\big(\bfC_k \bfA_k)\big|^2. 
\end{align*}
Since $\bfh_{\dm,k}, \:\bfT$ and $\bfr_{k}$ are mutually independent, and denoting $\channelT$ by $\bfT_r$, we have
\begin{align}    
\mathop{{}\mathbb{E}}\big[{\bfh_k^{\Hm}}\bfA_k{\bfh_k}{\bfh_k^{\Hm}}\bfA_k^{\Hm}\bfh_k \big] &= \mathop{{}\mathbb{E}}\big[|\bfh_{\dm,k}^{\Hm}\bfA_k\bfh_{\dm,k}|^2 \big] \nonumber \\
&+ \Ex\big[\bfh_{\dm,k}^{\Hm}\bfA_k\bfh_{\dm,k}\bfr_k^{\Hm}\Phimat^{\Hm}{\bfT} \bfA_k^{\Hm}{\bfT}^{\Hm}\Phimat\bfr_k\big] \nonumber \\
 &+ \Ex[|\bfh_{\dm,k}^{\Hm}\bfA_k \bfT^{\Hm}\Phimat\bfr_k|^2] \nonumber \\ & + \Ex[|\bfr_k^{\Hm}\Phimat^{\Hm}\bfT \bfA_k \bfh_{\dm,k}|^2] \nonumber \\ &+ \Ex[\bfr_k^{\Hm}\Phimat^{\Hm}\bfT \bfA_k\bfT^{\Hm}\Phimat\bfr_k\bfh_{\dm,k}^{\Hm}\bfA_k^{\Hm}\bfh_{\dm,k}] \nonumber \\
&+ \Ex[|\bfr_k^{\Hm}\Phimat^{\Hm}\bfT \bfA_k\bfT^{\Hm}\Phimat\bfr_k|^2 \big]. \label{eqn6}
\end{align}
For the computation of the terms in \eqref{eqn6}, we need some additional results.
\begin{align}    \mathop{{}\mathbb{E}}\big[\bfT_r \bfA_k \bfT_r^{\Hm} \big] &= \beta \mathop{{}\mathbb{E}} \left[\rirs^{1/2} {\bfW} \rtx^{1/2, \Hm} \bfA_k \rtx^{1/2} \bfW^{\Hm} \rirs^{1/2, \Hm}  \right] \nonumber \\
& = \beta \tr(\rtx \bfA_k) \rirs. \label{eqn7}
\end{align}
Using Lemma~2 of \cite{bjornson2015massive}, we get
\begin{align}
 &\mathop{{}\mathbb{E}}\big[ |  \bfr_k^{\Hm}\Phimat^{\Hm}\bfT_r\bfA_k\bfT_r^{\Hm}\Phimat\bfr_k|^2\big] \nonumber \\ & = \mathop{{}\mathbb{E}_{\bfT_r}}\left[ \left|  \tr(\bfC_{\bfr,k}\Phimat^{\Hm}\bfT_r \bfA_k\bfT_r^{\Hm}\Phimat)\right|^2\right] \nonumber\\
 &+ \mathop{{}\mathbb{E}_{\bfT_r}}\left[\tr \left( \bfC_{\bfr,k}\Phimat^{\Hm}\bfT_r \bfA_k\bfT_r^{\Hm}\Phimat\bfC_{\bfr,k}\Phimat^{\Hm}\bfT_r \bfA_k^{\Hm}\bfT_r^{\Hm}\Phimat \right)\right]. \label{app25} 
\end{align}
The first term in \eqref{app25} can be written as
\begin{align}   & \mathop{{}\mathbb{E}_{\bfT_r}}\left[ \left|  \tr(\bfC_{\bfr,k}\Phimat^{\Hm}\bfT_r \bfA_k\bfT_r^{\Hm}\Phimat)\right|^2 \right] \nonumber \\ %&= \beta^2  \mathop{{}\mathbb{E}}\left[ \left| \tr(\bfC_{\bfr,k}\Phimat^{\Hm}\rirs^{1/2} \bfW \rtx^{1/2, \Hm}\bfA_k \rtx^{1/2} \bfW^{\Hm} \rirs^{1/2, \Hm} \Phimat) \right|^2 \right] \nonumber \\
& = \mathop{{}\mathbb{E}_{\bfw}}\Bigg[ \Bigg| \bfw^{\Hm} \Bigg(\Ycal_{1,k}^{\Tm} \otimes \Big(\rirs^{1/2, \Hm}\Ycal_{2,k} \rirs^{1/2} \Big) \Bigg)\bfw   \Bigg|^2\Bigg]  
\end{align}
where $\Ycal_{1,k} = \rtx^{1/2, \Hm}\bfA_k \rtx^{1/2}$, $\Ycal_{2,k} = \beta^2\Phimat \bfC_{\bfr,k} \Phimat^{\Hm}$ and $\bfw = $ vec($\bfW$), and since the entries of $\bfW$ are i.i.d. with zero mean and unit variance, it follows that $\bfw \sim\nc({\bf{0}}, {\bf{I}})$. The above expression can hence be computed as follows
\begin{align}    &\mathop{{}\mathbb{E}_{\bfT_r}}\left[ \left|  \tr(\bfC_{\bfr,k}\Phimat^{\Hm}\bfT_r \bfA_k\bfT_r^{\Hm}\Phimat)\right|^2\right] \nonumber\\ &= \beta^2 \Bigg(\left|\tr\left(\rtx \bfA_k \right)\right|^2\left|\tr\left(\rirs\Phimat \bfC_{\bfr,k}\Phimat^{\Hm}\right) \right|^2 \nonumber \\
&+  \tr(\rtx \bfA_k^{\Hm}\rtx \bfA_k) \tr(\rirs \Phimat \bfC_{\bfr,k} \Phimat^{\Hm}\rirs \Phimat \bfC_{\bfr,k} \Phimat^{\Hm})\Bigg) \nonumber 
\end{align}
where we have used the property that for all square matrices $\bfP$, $\bfQ$, $\bfR$ and $\bfS$, we have $\tr\left(\bfP \otimes \bfQ\right) = \tr(\bfP) \tr(\bfQ)$, and $(\bfP \otimes \bfQ)(\bfR \otimes \bfS) = (\bfP \bfR)\otimes(\bfQ \bfS)$ \cite{brewer1978kronecker}. \\
\begin{align}
    &\mathop{{}\mathbb{E}_{\bfT_r}} \left[ \tr \left( \bfC_{\bfr,k}\Phimat^{\Hm}\bfT_r \bfA_k\bfT_r^{\Hm}\Phimat\bfC_{\bfr,k}\Phimat^{\Hm}\bfT_r \bfA_k^{\Hm}\bfT_r^{\Hm}\Phimat\right)\right]  \nonumber \\
   &= \beta^2 \mathop{{}\mathbb{E}_{\bfW}} \Bigg[ \tr \Biggl(\bfC_{\bfr,k}\Phimat^{\Hm}\rirs^{1/2} {\bfW} \rtx^{1/2, \Hm}\bfA_k\rtx^{1/2} \bfW^{\Hm} \rirs^{1/2, \Hm}  \nonumber \\
&\quad \times \Phimat\bfC_{\bfr,k}\Phimat^{\Hm}\rirs^{1/2} {\bfW} \rtx^{1/2, \Hm}\bfA_k^{\Hm}\rtx^{1/2} \bfW^{\Hm} \rirs^{1/2, \Hm}\Phimat\Biggr) \Bigg].  \nonumber  
\end{align}
Let the matrix $\bfC_{\bfr,k}^{1/2, \Hm}\Phimat^{\Hm} \rirs^{1/2}$ be decomposed by its SVD as $\bfU {\it{{\pmb{\Lambda}}}} \bfV^{\Hm}$, where $\bfU$ and $\bfV$ contain the singular vectors and ${\pmb{\Lambda}}$ is a diagonal matrix containing the singular values. Using the fact that $\bfV^{\Hm} {\bfW} = \bfZ$ has the same distribution as $\bfW$, since $\bfV$ is a unitary matrix, the above expression can be written as
\begin{align}
     \beta^2 \mathop{{}\mathbb{E}_{\bfZ}} \left[\tr \left({\pmb{\Lambda}}^2 \bfZ \Ycal_{1,k} \bfZ^{\Hm}  {\pmb{\Lambda}}^{2}   \bfZ \Ycal_{1,k}^{\Hm}  \bfZ^{\Hm} \right)\right]. \nonumber
\end{align}
Denoting the eigen values of $\bfC_{\bfr,k}^{1/2, \Hm}\Phimat^{\Hm} \rirs\Phimat \bfC_{\bfr,k}^{1/2}$ by $\lambda_i$s, the diagonal matrix ${\pmb{\Lambda}}$ can be written as $\sum\limits_{i}\lambda_i \bfe_i \bfe_i^{\Tm}$. Hence, the above expression is equivalent to 
\begin{align}
   & \beta^2 \sum\limits_{i,j}\lambda_i\lambda_j \mathop{{}\mathbb{E}} \Bigg[\tr \Bigg(\bfz_i \bfz_i^{\Hm}  \Ycal_{1,k}   \bfz_j \bfz_j^{\Hm} \Ycal_{1,k}^{\Hm} \Bigg)\Bigg] \nonumber
\end{align}
where the matrix $\bfZ^{\Hm}$ is denoted by $[\bfz_1, \cdots, \bfz_{\text{N}}]$. %The above expression can be further decomposed as 
%\begin{align}
%   & \beta^2 \sum_{\substack{i, j \\ i \neq j}}\lambda_i\lambda_j \mathop{{}\mathbb{E}} \left[\tr \left(\bfz_i \bfz_i^{\Hm}  \rtx^{1/2, \Hm}\bfA_k\rtx^{1/2}   \bfz_j \bfz_j^{\Hm} \rtx^{1/2, \Hm}\bfA_k^{\Hm}\rtx^{1/2}  \right)\right] \nonumber \\
 %   &+ \beta^2 \sum\limits_{i}\lambda_i^2\mathop{{}\mathbb{E}} \left[\tr \left(\bfz_i \bfz_i^{\Hm}  \rtx^{1/2, \Hm}\bfA_k\rtx^{1/2}   \bfz_i \bfz_i^{\Hm} \rtx^{1/2, \Hm}\bfA_k^{\Hm}\rtx^{1/2}  \right)\right] \nonumber
%\end{align}
Since $\bfz_i$s have the same distribution as $\bfw_i$s, $\bfz_{i}\sim\nc({\bf{0}}, {\bf{I}})$, and the expectation can be computed as
\begin{align}
   &\beta^2 \sum_{\substack{i, j \\ i \neq j}}\lambda_i\lambda_j \tr \left(  \rtx^{1/2, \Hm}\bfA_k\rtx^{1/2}  \rtx^{1/2, \Hm}\bfA_k^{\Hm}\rtx^{1/2}  \right) \nonumber \\ &+\beta^2 \sum\limits_{i}\lambda_i^2 \mathop{{}\mathbb{E}_{\bfz_i}} \left[\left|\bfz_i^{\Hm}  \rtx^{1/2, \Hm}\bfA_k\rtx^{1/2}   \bfz_i   \right|^2 \right] \nonumber \\
&\stackrel{(a)}= \beta^2 \sum_{\substack{i, j \\ i \neq j}}\lambda_i\lambda_j \tr \left(  \rtx^{1/2, \Hm}\bfA_k\rtx^{1/2}  \rtx^{1/2, \Hm}\bfA_k^{\Hm}\rtx^{1/2}  \right) \nonumber \\ &+ \beta^2 \sum\limits_{i}\lambda_i^2 \left( |\tr(\rtx \bfA_k)|^2 + \tr(\rtx \bfA_k \rtx \bfA_k^{\Hm}) \right) \nonumber \\
&= \beta^2 \sum\limits_{i, j }\lambda_i\lambda_j \tr \left(  \rtx\bfA_k\rtx \bfA_k^{\Hm}\right) + \beta^2 \sum\limits_{i}\lambda_i^2 \left|\tr(\rtx \bfA_k) \right|^2  \nonumber \\
&= \beta^2 \Big(\tr^2\left(\bfC_{\bfr,k}\Phimat^{\Hm}\rirs \Phimat \right) \tr \left( \rtx\bfA_k\rtx \bfA_k^{\Hm}\right) \nonumber \\ & + \tr\left(\bfC_{\bfr,k}\Phimat^{\Hm}\rirs \Phimat \bfC_{\bfr,k}\Phimat^{\Hm}\rirs \Phimat\right) \left|\tr(\rtx \bfA_k) \right|^2 \Big) \label{eqn10} 
\end{align}
where ($a$) again uses Lemma~2 of \cite{bjornson2015massive}. Now using \eqref{eqn7}-\eqref{eqn10}, Lemma~1 and Lemma~2 of \cite{bjornson2015massive}, all the terms in \eqref{eqn6} can be calculated in closed-form.
 Using the expression of $\bfC_k$ from \eqref{eqnC} and with some calculations, we get
  \begin{align}
     &\mathop{{}\mathbb{E}}\big[{\bfh_k^{\Hm}}\bfA_k{\bfh_k}{\bfh_k^{\Hm}}\bfA_k^{\Hm}\bfh_k\big]  =  \tr(\bfC_k\bfA_k\bfC_k\bfA_k^{\Hm}) + \left|\tr(\bfC_k\bfA_k)\right|^2  \nonumber \\ &   +  \beta^2 \tr\left(\bfQ\rirs \right) \times \left(\left|\tr(\rtx \bfA_k) \right|^2 + \tr(\rtx \bfA_k \rtx \bfA_k^{\Hm}) \right) \nonumber  \\ &+ 2 \:\beta \Re{\tr(\rtx \bfA_k^{\Hm})\tr(\bfQ \bar{\bfT}\bfA_k\bar{\bfT}^{\Hm})} \nonumber \\ & + \beta \tr(\bfQ \bar{\bfT}\bfA_k \rtx \bfA_k^{\Hm}\bar{\bfT}^{\Hm}) + \beta \tr(\bfQ \bar{\bfT}\bfA_k^{\Hm} \rtx \bfA_k\bar{\bfT}^{\Hm})
 \end{align}
 where $\bfQ = \Phimat \bfC_{\bfr,k}\Phimat^{\Hm}\rirs \Phimat \bfC_{\bfr,k}\Phimat^{\Hm}$.
\end{proof}

\subsection{Computation of the gradient}
From \eqref{eqnC}, we get
\begin{align}
\label{A.63}
 \bfc_k^{\Hm}\bfa_k = \tr\left(\bfC_k \bfA_k \right) & = \tr(\bfC_{\dm,k} \bfA_k) + \tr(\biggamma_k \bfV) \tr(\rtx \bfA_k) \nonumber \\ & + \tr(\Tmean^{\Hm}\Phimat\bfC_{\bfr,k}\Phimat^{\Hm}\Tmean\bfA_k)
\end{align}
where $\biggamma_k = \beta \left(\rirs \odot \bfC_{\bfr, k}^{\Tm} \right) $ and $\bfV = \phivec \phivec^{\Hm}$. Both $\biggamma_k$ and $\bfV$ are Hermitian matrices. The above expression can be further simplified as
\begin{align}
\label{A.64}
\begin{split}
 \bfc_k^{\Hm}\bfa_k &=  \tr(\bfC_{\dm,k} \bfA_k) + \tr\Bigg(\Tmean\bfA_k\Tmean^{\Hm}\left(\bfC_{\bfr,k}\odot \phivec\phivec^{\Hm}\right)\Bigg) \\ &+ \tr(\biggamma_k \bfV) \tr(\rtx \bfA_k)  \\
  &\stackrel{(a)}{=} \tr(\bfC_{\dm,k} \bfA_k) + \tr\Bigg(\Big(\big(\Tmean\bfA_k\Tmean^{\Hm}\big)\odot \bfC_{\bfr,k}^{\Tm}\Big)\bfV\Bigg) \\ &+ \tr(\biggamma_k \bfV) \tr(\rtx \bfA_k)  \\
  &= \tr(\bfC_{\dm,k} \bfA_k) + \tr(\Dcal_k \bfV) 
 \end{split}
\end{align}
where ($a$) follows from Lemma~1 in \cite{syed2023design} and $\Dcal_k$ is given by
\begin{align}
\label{A.65}
    \Dcal_k = \left(\Tmean\bfA_k\Tmean^{\Hm}\right)\odot \bfC_{\bfr,k}^{\Tm} + \tr(\rtx \bfA_k)\biggamma_k.
\end{align}
Similarly, we get
\begin{align}
\label{A.66}
\begin{split}
    \left|{{\bfc}_k^{\Hm}{\bfa}_k}\right|^2 &= %\tr\left(\bfC_k \bfA_k \right)\tr\left(\bfC_k \bfA_k^{\Hm} \right) \\
    \left| \tr(\bfC_{\dm,k} \bfA_k) \right|^2   + \tr(\Dcal_k \bfV)\tr\left(\bfC_{\dm, k} \bfA_k^{\Hm} \right)  \\ &+  \left| \tr(\Dcal_k \bfV)\right|^2 + \tr(\Dcal_k^{\Hm} \bfV)\tr\left(\bfC_{\dm, k} \bfA_k \right)
    \end{split}
\end{align}
\begin{align}
\label{A.67}
\begin{split}
    \bfa_j^{\Hm}\left({\bfC}_j^{\Tm} \otimes {\bfC}_k \right)\bfa_j 
   & = \tr({\bfC_{\dm, k} \bfA_j \bfC_j\bfA_j^{\Hm}})  \\ &+ \tr(\Tmean^{\Hm}\Phimat\bfC_{\bfr,k}\Phimat^{\Hm}\Tmean\bfA_j \bfC_j \bfA_j^{\Hm}) \\
   &+ \tr(\biggamma_k \bfV) \tr(\rtx \bfA_j \bfC_j \bfA_j^{\Hm}).  
    \end{split}
\end{align}
Further expanding the terms,
\begin{align}
    \tr({\bfC_{\dm, k} \bfA_j \bfC_j\bfA_j^{\Hm}})  & =   \tr({\bfC_{\dm, k} \bfA_j \bfC_{\dm, j}\bfA_j^{\Hm}})  \nonumber \\ &+ \tr({\bfC_{\dm, k} \bfA_j \Tmean^{\Hm}\Phimat\bfC_{\bfr,j}\Phimat^{\Hm}\Tmean\bfA_j^{\Hm}}) \nonumber \\ & + \tr(\biggamma_j \bfV) \tr({\bfC_{\dm, k} \bfA_j \rtx\bfA_j^{\Hm}})  
\end{align}
\begin{align}
\begin{split}    \tr &(\Tmean^{\Hm}\Phimat\bfC_{\bfr,k}\Phimat^{\Hm}\Tmean\bfA_j \bfC_j \bfA_j^{\Hm})  
 \\  &= \tr(\Tmean^{\Hm}\Phimat\bfC_{\bfr,k}\Phimat^{\Hm}\Tmean\bfA_j \bfC_{\dm, j} \bfA_j^{\Hm}) \\
    &+ \tr(\Tmean^{\Hm}\Phimat\bfC_{\bfr,k}\Phimat^{\Hm}\Tmean\bfA_j \Tmean^{\Hm}\Phimat\bfC_{\bfr,j}\Phimat^{\Hm}\Tmean \bfA_j^{\Hm}) \\
    &+ \tr(\biggamma_j \bfV)\tr(\Tmean^{\Hm}\Phimat\bfC_{\bfr,k}\Phimat^{\Hm}\Tmean\bfA_j \rtx \bfA_j^{\Hm})
    \end{split}
\end{align}
\begin{align}
\begin{split}
    \tr(\biggamma_k \bfV) \tr & (\rtx \bfA_j \bfC_j \bfA_j^{\Hm}) = \tr(\biggamma_k \bfV) \tr(\rtx \bfA_j \bfC_{\dm, j} \bfA_j^{\Hm}) \\ &+ \tr(\biggamma_k \bfV) \tr(\rtx \bfA_j \Tmean^{\Hm}\Phimat\bfC_{\bfr,j}\Phimat^{\Hm}\Tmean\bfA_j^{\Hm}) \\
    &+\tr(\biggamma_k \bfV)\tr(\biggamma_j \bfV) \tr(\rtx \bfA_j \rtx \bfA_j^{\Hm}).
    \end{split}
\end{align}
Denoting the constant terms which are independent of $\phivec$ as ${\ell}_1$, we get
\begin{align}
\label{A.73}
\begin{split}
     \bfa_j^{\Hm} & \left({\bfC}_j^{\Tm} \otimes {\bfC}_k \right) \bfa_j  =  \tr({\bfC_{\dm, k} \bfA_j \Tmean^{\Hm}\Phimat\bfC_{\bfr,j}\Phimat^{\Hm}\Tmean\bfA_j^{\Hm}}) \\ & + \tr(\biggamma_j \bfV) \tr({\bfC_{\dm, k} \bfA_j \rtx\bfA_j^{\Hm}}) \\ 
     &+  \tr(\Tmean^{\Hm}\Phimat\bfC_{\bfr,k}\Phimat^{\Hm}\Tmean\bfA_j \bfC_{\dm, j} \bfA_j^{\Hm}) \\
    &+ \tr(\Tmean^{\Hm}\Phimat\bfC_{\bfr,k}\Phimat^{\Hm}\Tmean\bfA_j \Tmean^{\Hm}\Phimat\bfC_{\bfr,j}\Phimat^{\Hm}\Tmean \bfA_j^{\Hm}) \\
    &+ \tr(\biggamma_j \bfV)\tr(\Tmean^{\Hm}\Phimat\bfC_{\bfr,k}\Phimat^{\Hm}\Tmean\bfA_j \rtx \bfA_j^{\Hm}) \\
    &+\tr(\biggamma_k \bfV) \tr(\rtx \bfA_j \bfC_{\dm, j} \bfA_j^{\Hm}) \\ &+ \tr(\biggamma_k \bfV) \tr(\rtx \bfA_j \Tmean^{\Hm}\Phimat\bfC_{\bfr,j}\Phimat^{\Hm}\Tmean\bfA_j^{\Hm}) \\
    &+\tr(\biggamma_k \bfV)\tr(\biggamma_j \bfV) \tr(\rtx \bfA_j \rtx \bfA_j^{\Hm}) + {\ell}_1
     \end{split}
\end{align}
where ${\ell}_1$ consists of the terms which are independent of $\phivec$. The terms can be further simplified as
\begin{align}
\label{part1}
\begin{split}
    \tr & ({\bfC_{\dm, k}  \bfA_j \Tmean^{\Hm}\Phimat\bfC_{\bfr,j}\Phimat^{\Hm}\Tmean\bfA_j^{\Hm}}) \\
  %& =  \tr\Bigg(\Tmean\bfA_j^{\Hm}\bfC_{\dm, k} \bfA_j \Tmean^{\Hm}\left(\bfC_{\bfr,j}\odot\phivec\phivec^{\Hm} \right)\Bigg)  \\
  & =  \tr\Bigg(\Big(\big(\Tmean\bfA_j^{\Hm}\bfC_{\dm, k} \bfA_j \Tmean^{\Hm}\big)\odot \bfC_{\bfr,j}^{\Tm}\Big)\phivec\phivec^{\Hm}\Bigg).
    \end{split}
\end{align}
Similarly, we have
\begin{align}    
&\tr \left(\Tmean^{\Hm}\Phimat\bfC_{\bfr,k}\Phimat^{\Hm}\Tmean\bfA_j \bfC_{\dm, j} \bfA_j^{\Hm} \right) \nonumber \\ & \quad \quad \quad \quad = \tr\Bigg(\Big(\big(\Tmean\bfA_j \bfC_{\dm, j} \bfA_j^{\Hm}\Tmean^{\Hm}\big)\odot \bfC_{\bfr,k}^{\Tm}\Big)\phivec\phivec^{\Hm}\Bigg) \\   &\tr \left(\Tmean^{\Hm}\Phimat\bfC_{\bfr,k}\Phimat^{\Hm}\Tmean\bfA_j \rtx \bfA_j^{\Hm} \right) \nonumber \\ & \quad \quad \quad \quad = \tr\Bigg(\Big(\big(\Tmean\bfA_j \rtx \bfA_j^{\Hm}\Tmean^{\Hm}\big)\odot\bfC_{\bfr,k}^{\Tm}\Big)\phivec\phivec^{\Hm}\Bigg) \\
  & \tr(\rtx \bfA_j \Tmean^{\Hm}\Phimat\bfC_{\bfr,j}\Phimat^{\Hm}\Tmean\bfA_j^{\Hm}) \nonumber \\ & \quad \quad \quad \quad  = \tr\Bigg(\Big(\big(\Tmean\bfA_j^{\Hm} \rtx \bfA_j\Tmean^{\Hm}\big)\odot\bfC_{\bfr,j}^{\Tm}\Big)\phivec\phivec^{\Hm}\Bigg). \label{part2}
\end{align}
Now, the first term in \eqref{6.20} can be simplified using \eqref{A.64} as follows
\begin{align}
    \sum\limits_{k=1}^{K} 2 & \left(\sqrt{1 + \bar{\lambda}_k}\right) \text{Re}\left\{\bar{\chi}_k^{*}{\bfc}_k^{\Hm}\bar{\bfa}_k\right\}  \nonumber \\ &=  \sum\limits_{k=1}^{K} 2\left(\sqrt{1 + \bar{\lambda}_k}\right)\text{Re}\left\{\bar{\chi}_k^{*}\tr(\bar{\Dcal}_k \bfV) \right\} + \ell_2 \nonumber\\
& = \phivec^{\Hm} \Xcal_1 \phivec + \ell_2
\end{align}
where $\ell_2$ consists of the terms which are independent of $\phivec$, and $\bar{\Dcal}_k$ and $\Xcal_1$ are given by
\begin{align}
\bar{\Dcal}_k & = \left(\Tmean\bar{\bfA}_k\Tmean^{\Hm}\right)\odot \bfC_{\bfr,k}^{\Tm} + \tr(\rtx \bar{\bfA}_k)\biggamma_k \\
    \Xcal_1 & =  \sum\limits_{k=1}^{K} \left(\sqrt{1 + \bar{\lambda}_k}\right)\left(\bar{\chi}_k^{*}\bar{\Dcal}_k + \bar{\chi}_k\bar{\Dcal}_k^{\Hm}  \right).
\end{align}
The second term in \eqref{6.20} can be simplified using \eqref{A.66}. 
\begin{align}
\begin{split}
    \sum\limits_{k=1}^{K}\left|\bar{\chi}_k\right|^2\left|{{\bfc}_k^{\Hm}\bar{\bfa}_k}\right|^2 & = \sum\limits_{k=1}^{K} \left| \phivec^{\Hm} \Xcal_{2,k} \phivec \right|^2 + \phivec^{\Hm} \Xcal_3 \phivec + \ell_3
    \end{split}
\end{align}
where $\ell_3$ denotes the constant terms which are irrelevant for the optimisation, and
\begin{align}
    &\Xcal_{2,k}  = \left|\bar{\chi}_k\right| \bar{\Dcal}_k \\
    &\Xcal_3  =  \sum\limits_{k=1}^{K}\left|\bar{\chi}_k\right|^2 \left(\tr\left(\bfC_{\dm, k} \bar{\bfA}_k^{\Hm} \right)\bar{\Dcal}_k + \tr\left(\bfC_{\dm, k} \bar{\bfA}_k \right) \bar{\Dcal}_k^{\Hm} \right).
\end{align}
The third term in \eqref{6.20} can be simplified using \eqref{A.73} - \eqref{part2}. 
\begin{align}
\begin{split}    
& \sum\limits_{k=1}^{K}\left|\bar{\chi}_k\right|^2\sum\limits_{j=1}^{K}{\bar{\bfa}_j^{\Hm}\left({\bfC}_j^{\Tm} \otimes {\bfC}_k \right) \bar{\bfa}_j} \\
    & = \sum\limits_{k=1}^{K}\left|\bar{\chi}_k\right|^2 \phivec^{\Hm} \Xcal_{4,k} \phivec + \sum\limits_{k=1}^{K} \left(\phivec^{\Hm}\biggamma_k \phivec \right)\left( \phivec^{\Hm}  \Xcal_{5,k} \phivec\right) \\
    &+  \left(\phivec^{\Hm}\Big(\sum\limits_{k=1}^{K}\left|\bar{\chi}_k\right|^2\biggamma_k \Big)\phivec \right)\Bigg( \phivec^{\Hm} \Xcal_{6} \phivec\Bigg) \\
    &+ \sum\limits_{j=1}^{K} \tr\Bigg( \bfS_j^{\Hm}\Phimat\Big(\sum\limits_{k=1}^{K}\left|\bar{\chi}_k\right|^2\bfC_{\bfr,k}\Big)\Phimat^{\Hm}\bfS_j\Phimat \bfC_{\bfr,j}\Phimat^{\Hm}\Bigg) + \ell_4
    \end{split}
\end{align}
where $\ell_4$ denotes the constant terms, which are independent of $\phivec$, and
\begin{align}
    \Xcal_{4,k} & =  \sum\limits_{j=1}^{K}\left(\Tmean\bar{\bfA}_j^{\Hm}\bfC_{\dm,k}\bar{\bfA}_j\Tmean^{\Hm}\right)\odot \bfC_{\bfr,j}^{\Tm} \nonumber \\ &+  \sum\limits_{j=1}^{K}\tr({\bfC_{\dm, k} \bar{\bfA}_j \rtx\bar{\bfA}_j^{\Hm}})\biggamma_j \nonumber \\
    &+ \sum\limits_{j=1}^{K}\left(\Tmean\bar{\bfA}_j\bfC_{\dm,j}\bar{\bfA}_j^{\Hm}\Tmean^{\Hm}\right)\odot \bfC_{\bfr,k}^{\Tm} \nonumber \\ &+  \sum\limits_{j=1}^{K} \tr(\rtx \bar{\bfA}_j \bfC_{\dm, j} \bar{\bfA}_j^{\Hm})\biggamma_k     \\
    \Xcal_{5,j} & =  \left(\Tmean\bar{\bfA}_j\rtx\bar{\bfA}_j^{\Hm}\Tmean^{\Hm}\right)\odot \left(\sum\limits_{k=1}^{K} \left|\bar{\chi}_k\right|^2\bfC_{\bfr,k}^{\Tm} \right) \\
    \Xcal_6 & = \sum\limits_{j=1}^{K}\left(\Tmean\bar{\bfA}_j^{\Hm}\rtx\bar{\bfA}_j\Tmean^{\Hm}\right)\odot \bfC_{\bfr,j}^{\Tm} \nonumber \\ & +  \sum\limits_{j=1}^{K}\tr(\rtx \bar{\bfA}_j \rtx \bar{\bfA}_j^{\Hm})\biggamma_j \\
    \bfS_j &= \Tmean \bfA_j \Tmean^{\Hm}.
\end{align}
Next, $\bfa_k^{\Hm}\bfJ_k\bfa_k$ can be written as
\begin{align}
 % & \bfa_k^{\Hm}\bfJ_k\bfa_k =   \beta^2 \tr\left(\bfC_{\bfr,k}\Phimat^{\Hm}\rirs \Phimat \bfC_{\bfr,k}\Phimat^{\Hm}\rirs \Phimat\right) \nonumber \\ & \quad \quad \quad \times \left(\left|\tr(\rtx \bfA_k) \right|^2 + \tr(\rtx \bfA_k \rtx \bfA_k^{\Hm}) \right) \nonumber \\
 %     &+ \beta \tr(\rtx \bfA_k^{\Hm})\tr(\bfC_{\bfr,k}\Phimat^{\Hm}\bar{\bfT}\bfA_k\bar{\bfT}^{\Hm}\Phimat \bfC_{\bfr,k}\Phimat^{\Hm}\rirs \Phimat) \nonumber \\
 %     &+ \beta \tr(\rtx \bfA_k)\tr(\bfC_{\bfr,k}\Phimat^{\Hm}\bar{\bfT}\bfA_k^{\Hm}\bar{\bfT}^{\Hm}\Phimat \bfC_{\bfr,k}\Phimat^{\Hm}\rirs \Phimat) \nonumber \\
 %     & + \beta \tr(\bfC_{\bfr,k}\Phimat^{\Hm}\rirs \Phimat \bfC_{\bfr,k}\Phimat^{\Hm}\bar{\bfT}\bfA_k\rtx\bfA_k^{\Hm}\bar{\bfT}^{\Hm}\Phimat) \nonumber \\
 %     & + \beta \tr(\bfC_{\bfr,k}\Phimat^{\Hm}\rirs \Phimat \bfC_{\bfr,k}\Phimat^{\Hm}\bar{\bfT}\bfA_k^{\Hm}\rtx\bfA_k\bar{\bfT}^{\Hm}\Phimat) \nonumber \\
   \bfa_k^{\Hm}\bfJ_k\bfa_k   &= \tr\left(\bfC_{\bfr,k}\Phimat^{\Hm}\bfM_k \Phimat \bfC_{\bfr,k}\Phimat^{\Hm}\rirs \Phimat\right)
\end{align}
where
\begin{align*}
    \bfM_k &= \beta^2 \left(\left|\tr(\rtx \bfA_k) \right|^2 + \tr(\rtx \bfA_k \rtx \bfA_k^{\Hm}) \right)\rirs  \\ &+ \beta \tr(\rtx \bfA_k^{\Hm})\bar{\bfT}\bfA_k\bar{\bfT}^{\Hm} +  \beta \tr(\rtx \bfA_k)\bar{\bfT}\bfA_k^{\Hm}\bar{\bfT}^{\Hm}\\
    &+ \beta \bar{\bfT}\bfA_k\rtx\bfA_k^{\Hm}\bar{\bfT}^{\Hm} +  \beta \bar{\bfT}\bfA_k^{\Hm}\rtx\bfA_k\bar{\bfT}^{\Hm}.
\end{align*}
Finally, ignoring all the constant terms, \eqref{6.20} can be written as a function of $\phivec$ as
\begin{align}
\label{A.84}
  & f(\phivec)   =  \phivec^{\Hm} \Gcal \phivec - \sum\limits_{k=1}^{K} \left| \phivec^{\Hm} \Xcal_{2,k} \phivec \right|^2 - \sum\limits_{k=1}^{K} \left(\phivec^{\Hm}\biggamma_k \phivec \right)\left( \phivec^{\Hm}  \Xcal_{5,k} \phivec\right)  \nonumber \\
    &-  \left(\phivec^{\Hm}\Big(\sum\limits_{k=1}^{K}\left|\bar{\chi}_k\right|^2\biggamma_k \Big)\phivec \right)\Bigg( \phivec^{\Hm}  \Xcal_{6} \phivec\Bigg)   \nonumber \\
    &- \sum\limits_{j=1}^{K} \tr\Bigg( \bfS_j^{\Hm}\Phimat\Big(\sum\limits_{k=1}^{K}\left|\bar{\chi}_k\right|^2\bfC_{\bfr,k}\Big)\Phimat^{\Hm}\bfS_j\Phimat \nonumber \bfC_{\bfr,j}\Phimat^{\Hm}\Bigg)  \nonumber \\
    &- \sum\limits_{k=1}^{K} \left|\bar{\chi}_k\right|^2\tr\left(\bfC_{\bfr,k}\Phimat^{\Hm}\bfM_k \Phimat \bfC_{\bfr,k}\Phimat^{\Hm}\rirs \Phimat\right)
\end{align}
where $\Gcal$ is given by
\begin{align}
    \Gcal = \Xcal_1 - \Xcal_3 -\sum\limits_{k=1}^{K} \left|\bar{\chi}_k\right|^2\Xcal_{4,k}.
\end{align}
The gradient of $f(\phivec)$ in \eqref{A.84} w.r.t. $\phivec^{*}$ is given by $\boldsymbol{\Delta}$, which can be expressed as follows 
\begin{align}
\label{A.86}
\begin{split}
   &\boldsymbol{\Delta} = \Gcal \phivec -  \sum\limits_{k=1}^{K} \left( \phivec^{\Hm} \Xcal_{2,k} \phivec \right) \Xcal_{2,k}^{\Hm} \phivec - \sum\limits_{k=1}^{K} \left( \phivec^{\Hm} \Xcal_{2,k}^{\Hm} \phivec \right) \Xcal_{2,k} \phivec \\ &- \sum\limits_{k=1}^{K} \left(\phivec^{\Hm}\biggamma_k \phivec \right)\Xcal_{5,k} \phivec 
    - \sum\limits_{k=1}^{K} \left(\phivec^{\Hm}\Xcal_{5,k} \phivec \right)\biggamma_k \phivec \\ &-  \left(\phivec^{\Hm}\bfY_1\phivec \right) \Xcal_{6} \phivec - \left(\phivec^{\Hm} \Xcal_{6} \phivec \right)  \bfY_1\phivec 
   \\ &- \sum\limits_{k=1}^{K} \left(\bfY_{3,k}\odot \bfC_{\bfr, k}^{\Tm} \right)\phivec - \sum\limits_{k=1}^{K} \left(\bfY_{4,k}\odot \bfY_2^{\Tm}\right)\phivec \\
    &- \sum\limits_{k=1}^{K} \left|\bar{\chi}_k\right|^2\left(\bfY_{5,k}\odot \bfC_{\bfr,k}^{\Tm}\right)\phivec- \sum\limits_{k=1}^{K} \left|\bar{\chi}_k\right|^2\left(\bfY_{5,k}^{\Hm}\odot \bfC_{\bfr,k}^{\Tm}\right)\phivec
    \end{split}
\end{align}
where
\begin{align*}
    \bfY_1  &= \sum\limits_{k=1}^{K}\left|\bar{\chi}_k\right|^2\biggamma_k, \bfY_2  = \sum\limits_{k=1}^{K}\left|\bar{\chi}_k\right|^2\bfC_{\bfr,k} , \\\bfY_{3, k}  &=  \bfS_k^{\Hm}\Phimat \bfY_2\Phimat^{\Hm}\bfS_k, \bfY_{4, k} =   \bfS_k\Phimat \bfC_{\bfr,k}\Phimat^{\Hm}\bfS_k^{\Hm}, \\ \bfY_{5, k} &=  \rirs \Phimat  \bfC_{\bfr,k}\Phimat^{\Hm} \bfM_k.
\end{align*}

\bibliographystyle{IEEEtran}
\bibliography{bibliography}
\end{document}